\documentclass[11pt]{article}

\usepackage[greek,american]{babel}
\usepackage[iso-8859-7]{inputenc}
\usepackage{amsmath,amsfonts,amssymb,latexsym}

\newtheorem{theorem}{Theorem}[section]
\newtheorem{definition}{Definition}[section]
\newtheorem{lemma}[theorem]{Lemma}
\newtheorem{example}[theorem]{Example}

\newtheorem{proposition}[theorem]{Proposition}

\newenvironment{proof}[1][Proof]{\textsc{#1.} }{\ \rule{0.5em}{0.5em}}
\numberwithin{equation}{section}

\begin{document}
\title{\Huge{COSMOLOGICAL SINGULARITIES}\thanks{To be published in the Springer
LNP Proceedings of the First Aegean Summer School of Cosmology
held on Samos, Greece, in September 21-29, 2001.}}
\author{{\Large Spiros Cotsakis}\\
\emph{Research Laboratory of Geometry, Dynamical Systems and
Cosmology}\\ \emph{Department of Mathematics, University of the
Aegean}\\ \emph{Karlovassi 83 200, Samos, Greece}\\ \texttt{email:
skot@aegean.gr}} \maketitle
\begin{abstract}
\noindent An overview is provided of the singularity theorems in
cosmological contexts at a level suitable for advanced graduate
students. The necessary background from tensor and causal geometry
to understand the theorems is supplied, the mathematical notion of
a cosmology is described in some detail and issues related to the
range of validity of general relativity are also discussed.
\end{abstract}
\newpage
\tableofcontents
\newpage
\section{Introduction}\label{cotsakis: Introduction}
General relativity is the best theory we have for a dynamical
description of spacetime, matter and gravitation. One might once
hoped, with Einstein, that the evolution and structure of
spacetime according to this theory would be free of singularities
-- places where the validity and predictions of the theory break
down -- and hence general relativity would represent the `final'
theory for the description of the physical world at the
macroscopic level. However, general relativity assumes and implies
the existence of spacetime and so the question of how such a
spacetime structure can be created is logically outside the realm
of this theory. This would not be a valid scientific question to
ask if general relativity had an infinite range of validity or,
equivalently, was a theory free from spacetime singularities for
in that case it would have represented the final physical theory
for macroscopic phenomena.

Unfortunately, the singularity theorems,  first proven more than
30 years ago by Stephen Hawking, Robert Geroch and Roger Penrose,
provide us with the bad news that indeed under certain conditions
all generic spacetimes of general relativity will disappear in
spacetime singularities either in the future or in the past. If
this is true and the mathematical structure of general relativity
implies the existence of spacetime singularities, then this theory
cannot be the final one for the consistent description of the
world. Its range of validity in such a case is finite and we must
look for another theory that will answer the, now valid, question
of how the spacetime of general relativity was created.

In such a new theory, the assumptions of the singularity theorems
will loose their meaning in much the same way as that in which the
assumptions of the Pythagorean theorem become vacuous in a curved
space. This unknown, new, fully consistent framework has currently
two offshoots: The first goes by the name of \emph{Quantum Theory
of Gravity} and asks for the complete connection of quantum
mechanics and general relativity. It is hoped that in such a
theory singularities are smoothed out in some way and a meaning of
how the universe begun will emerge. The currently popular general
approach to this problem is that quantum mechanics stays
`untouched' but general relativity is the one that needs
modification. If this is assumed then there are four different
ways in which modifications of general relativity can be
accomplished:
\begin{enumerate}
  \item Modify the action for gravity for example through the higher-order or
  scalar-tensor actions discussed below
  \item Increase the number of spacetime dimensions in the `old
  Kaluza-Klein' or in the new `Brane approach' way
  \item Introduce supersymmetry in the spacetime coordinates
  \item Via String theory effective actions in which the spacetime worldlines are
  replaced by higher dimensional string worldsheets usually in
  combination with the three approaches above.
\end{enumerate}
It is not our purpose here to discuss the different opinions which
exist for these issues but only to state that there \emph{are} at
least two other, distinct from 1-4 above, quantum gravity
approaches: One is the so-called \emph{Euclidean Quantum Gravity}
approach of Hawking and collaborators utilizing a complex-time
approach through path integration, an offshoot of which is the
theory of Quantum Cosmology (see for instance the collection of
reprint papers in \cite{gi-ha93}). Last but not least, we mention
an approach that Penrose has developed over the years and takes
the point of view that \emph{it is not general relativity but
quantum mechanics that needs modification and this modification
will come as a consequence of our physical theory of spacetime --
general relativity} (see \cite{pe89} for a popular account of this
interesting set of ideas). It is also true that many important
unsolved questions remain in this field.

The second, completely different (and inequivalent to the above)
set of ideas in the search for the unknown, new theory is
described by the \emph{String Theory/Noncommutative Geometry}
interface. In this framework, general relativity is not to be
quantized but the problem is how classical spacetime, and general
relativity, \emph{emerges} from a completely new theory in which
spacetime does not exist. What seem to exist instead at this level
are some algebraic operators and certain abstract
algebraic-geometric relations between them. It is not our purpose
here to enter into the very interesting and, by and large, open
problems of this highly complicated subject which is now in a
state of intense development (cf. \cite{ncg,st}).

Of course an important issue not yet fully decided is connected
with the \emph{nature} of spacetime singularities in general
relativity. All attempts up to now in this direction show that
indeed singularities are typically connected with very complicated
structures leading to the view discussed above. It may, however,
be that spacetime singularities in general relativity eventually
resemble the hydrodynamical shocks, but real progress in this
direction seems almost impossible at present. We may therefore
conclude that it is interesting (and legitimate!) to work on both
aspects of the problem, namely, search for a new framework which
will give meaning to spacetime itself or establish the true nature
of generic spacetime singularities in general relativity.

Finalizing this speculative Introduction, we mention a last
ingredient in our research path towards understanding these issues
-- cosmic censorship. This conjecture says that generic spacetimes
do not develop any singularities \emph{which are visible from
infinity}. That is, if this is true we may regard general
relativistic singularities even if they exist in a most complex
form as not so bad  features after all and continue to believe in
the `eternal power' of general relativity. For a readable account
of cosmic censorship, we refer the reader to \cite{pe79}.

It is therefore very basic to understand what is meant precisely
by a singularity in general relativity and under what conditions
singularities are expected to arise.  We provide below an
introduction to the singularity theorems of general relativity
with special emphasis to those theorems that predict, under
certain assumptions, the existence of spacetime singularities in
the cosmological context. In the next Section we show what a
cosmology is mathematically and how the results of the singularity
theorems, usually assumed to hold only for general relativity, are
in fact true for all theories which are conformally equivalent to
general relativity provided the other assumptions of the theorems
hold. More importantly, we show that violations of these theorems
in such theories represent special cases and are not generic
violations for these theories can all be regarded as containing
special splitted forms of the energy-momentum tensors as compared
to that of general relativity. Sections 3-7 provide the necessary
background in spacetime geometry to understand the proof of the
simplest singularity theorem and the statement of the most general
result of this type given in Section 8. Finally, Section 9 gives
(an introduction to) the physical interpretation of cosmological
applications of the spacetime geometry techniques developed
previously.

\section{Cosmologies}\label{cosmologies}
The basic object of study in any mathematical approach to
cosmological problems is that of \emph{a cosmology}.  There are
three essential elements that go into a cosmology:
\begin{itemize}
\item  A cosmological spacetime \texttt{(CS)}
\item  A theory of gravity \texttt{(TG)}
\item  A collection of matterfields \texttt{(MF)}.
\end{itemize}
A cosmology is a particular way of combining these three basic
elements into a meaningful whole:
\begin{equation}
\mathtt{Cosmology = CS+TG+MF}. \label{1}
\end{equation}
Examples of cosmologies can be constructed by taking entries from
the following table:
\begin{table}[ht]
\begin{center}
\begin{tabular}{l|c|r}
\multicolumn{3}{c}{\emph{\textbf{Cosmologies}}}\cr\hline {\bf
Theories of gravity} & \textbf{Cosmological spacetimes} &
\textbf{Matterfields}\cr\hline\hline General Relativity &
Isotropic & Vacuum\cr Higher Derivative Gravity & Homogeneous &
Fluids \cr Scalar-Tensor theories & Inhomogeneous & Scalarfields
\cr String theories & Generic & $n$-form fields \cr\hline
\end{tabular}
\end{center}\label{table1}
\caption{{\small Several members of each particular element of the
three essential elements comprising a cosmology according  Eq.
(\ref{1}).}}
\end{table}

For instance we can consider the families:
\begin{itemize}
\item  FRW/GR/vacuum
\item  Bianchi/ST/fluid
\item  Inhomogeneous/String/$n$-form
\item  Generic/GR/vacuum
\end{itemize}
\noindent and so on. In other Courses of this School you will
study mathematical and physical properties of  these and other
cosmologies using exact (i.e., closed form) or special
cosmological solutions. In this Course, however, we are interested
in cosmological applications of certain geometric methods and
results valid \emph{independently} of specific choices of
spacetimes, theories of gravity and matterfields, that is we
develop (and then apply) \emph{generic and global methods}.

The different theories of gravity appearing above are described by
\emph{actions} and their variations through the action principle
give their associated field equations. For instance general
relativity is described by the Einstein-Hilbert action and
associated Einstein field equations:
\begin{equation}\label{gr}
  S=\int_{\mathcal{M}}\bigl( R+L_{m}\bigr)dv_{g}\;\Rightarrow\;
  R_{ab}-\frac{1}{2}\;g_{ab}R=T_{ab}.
\end{equation}
Here $(\mathcal{M},g)$ is the spacetime manifold with metric
tensor $g_{ab},\;$ $dv_{g}$ is the invariant volume element of
$\mathcal{M},\; R_{ab}$ is the Ricci curvature tensor, $R$ the
scalar curvature and $T_{ab}$ is the stress-energy tensor of the
various matterfields appearing in the action defined through the
\emph{matter term} $L_{m}$  by
\begin{equation}\label{stress}
  T_{ab}=-\Bigl( \frac{\partial L_{m}}{\partial
  g^{ab}}-g_{ab}L_{m}\Bigr).
\end{equation}
A general action for higher order gravity theories can be taken to
be,
\begin{equation}\label{hog}
   S=\int_{\mathcal{M}}\bigl( f(R)+L_{m}\bigr) dv_{g},
\end{equation}
where $f(R)$ is an arbitrary smooth function of the scalar
curvature $R$. Variation of this action leads to higher order
field equations for the metric tensor describing the gravitational
field. Next scalar-tensor theories are described by the general
action,
\begin{equation}\label{s-t}
   S=\int_{\mathcal{M}}\bigl( f(\phi ,R)-g^{ab}\partial_{a}\phi\;\partial_{b}\phi
   +L_{m}\bigr)dv_{g},
\end{equation}
where we see the general coupling of the scalar field $\phi$ to
the curvature. Furthermore string theories are multi-dimensional,
scalar-tensor theories having additional scalar fields, formfields
and supersymmetry.

Hence there is a large variety of seemingly different actions for
the description of the gravitational field. A natural question
arises as to which of these actions (if any) describes gravity
correctly at all scales. In this connection, we quote a theorem
proved in \cite{ba-co88,ma89,co-mi01} which clarifies the
conformal structure of these (and other) gravity actions and shows
in addition that all couplings of the scalar field to the kinetic
term and to the curvature are essentially equivalent.
\begin{theorem}[Conformal equivalence]
All higher order, scalar-tensor and string actions are conformally
equivalent to general relativity with additional scalar fields
which have particular (different in each case) self-interaction
potentials.
\end{theorem}
Hence all these actions are in fact \emph{special cases} of the
general relativity action (\ref{gr}) in the sense that any
prediction made using any of the above variants of general
relativity can also be made by going to the corresponding
conformally related Einstein-matter system satisfied by the
conformally equivalent metric tensor,
\begin{equation}\label{conformal trans}
  \tilde{g}_{ab}=\Omega^{2}g_{ab},\; \Omega^{2}>0.
\end{equation}
Here $\Omega^{2}$ is a positive function of the fields and takes a
particular form when one conformally transforms a given action.
For instance one takes $\Omega^{2}=f'(R)$ for higher order gravity
whereas $\Omega^{2}=\phi$ for the simplest scalar-tensor theory,
Brans-Dicke gravity. For these reasons we `restrict' our attention
to general relativity or, more accurately, the Einstein-Hilbert
action (\ref{gr}).

It is indeed true that the conclusions of the singularity theorems
when applied to cosmology can be changed when one modifies the
action for gravity (for instance as in string theory) or considers
`matter' satisfying some `exotic' energy condition etc. As a
consequence of the above theorem on conformal equivalence,
however, our viewpoint towards these changes is that they
represent \emph{special cases} and are not generic. In all such
cases the matter lagrangian $\widetilde{L}_{m}$ in the so-called
Einstein frame (and so the stress tensor) as a rule naturally
\emph{splits} into a term describing the lagrangian of one or more
scalar fields, $\widetilde{L}_{\phi}$, and another term
$\widetilde{L}_{M}$ containing the remaining matter terms (fluids,
formfields, vacuum, etc), i.e.,
\begin{equation}\label{split stress}
  \widetilde{L}_{m}=\widetilde{L}_{\phi}+\widetilde{L}_{M},
\end{equation}
something that is not necessary in general. The singularity
theorems, in particular, do \emph{not} require such a splitting
for the matter tensor among their assumptions. We therefore see
that, as a consequence of the conformal equivalence theorem, all
choices above (and others related to those) for possible
`different' gravity theories are actually included under the
umbrella of the general Einstein-Hilbert action.

We return to our basic theme. There are two main areas in the
study of global methods -- \emph{singularities and evolution}. We
shall provide here an introduction to the theory of spacetime
singularities and their applications in cosmology. The chapter by
Choquet-Bruhat and York in this Volume, introduces the other main
area in generic methods -- global evolution.

A basic premise of this whole approach of applying spacetime
differential geometry methods to cosmology and gravitation is that
gravity is not like other physical fields. It is completely
different because it shapes the spacetime on which it acts, while
other fields act on a fixed background spacetime. In many respects
we find ourselves in a similar situation as that of the ancient
Greek geometers who, having constructed a  fully consistent
mathematical framework -- euclidean geometry, proved many theorems
corresponding to physical situations that gave rise to many
interesting models of physical reality and eventually to the new
framework we have today. Spacetime geometry and general relativity
represent a radical departure from euclidean geometry that is here
to stay irrespective of possible limitations (brought by the
singularity theorems discussed below) that, one hopes, will lead
in time to a new geometry and its physical interpretation.

The reader is referred to \cite{co01} for an introduction to
current cosmological issues without the use of mathematics.
Spacetime differential geometry, of which singularity theory forms
an important component, has only comparatively recently (with a
few important exceptions) stimulated the interest of
mathematicians. It is really quite distinct from the more common
Riemannian geometry, the latter now occurring in the study of
submanifolds representing an instant of time in spacetime.
Spacetime geometry has three branches, namely, the standard
\emph{tensor geometry} wherein curvature and geodesics are
described, the more recent \emph{causal geometry} where the
various causality properties of spacetimes belong and \emph{spin
geometry} which is that branch concerned with spinor objects
generalizing the usual tensor approach to geometry (or the
equivalent exterior differential forms). Singularity theory, which
we describe in these notes, can be thought of as belonging to the
interface of tensor and causal spacetime geometry. Helpful,
general sources on spacetime geometry which are recommended for
further reading are \cite{mtw73}-\cite{ch-de00}. General
references for cosmological applications of the material in these
notes are \cite{ha-el73}, \cite{wa84} and \cite{kr99}.

\section{The spacetime metric}
We assume some knowledge of manifolds, tensors and forms. An
$n$-\emph{manifold} $\mathcal{M}$ is a topological space that
locally looks like the Euclidean space $\mathbb{R}^{n}.$ This
means that there is a homeomorphism (usually called a coordinate
system or chart) of an open set of $\mathcal{M}$ onto an open set
of $\mathbb{R}^{n}.$ We assume that all manifolds in these notes
are 4-dimensional, but actually all results will be valid in
$n\geq2$ dimensions. Also our manifolds will be Hausdorff and
connected.

Now let $x\in\mathcal{M}$ and denote by $T_{x}\mathcal{M}$ the
tangent space of $\mathcal{M}$ at $x,$ $T_{x}\mathcal{M}^*$ the
corresponding cotangent space, $X(\mathcal{M})$  the set of all
smooth vectorfields on $\mathcal{M}$ and $F(\mathcal{M})$ the set
of all smooth, real--valued functions on $\mathcal{M}$. For any
non--negative integers $r,s,$ \emph{a tensor of type} $(r,s)$
\emph{at} $x$ is an $\mathbb{R}-$multilinear map $A:$
$(T_{x}\mathcal{M}^{\ast})^{r}\times(T_{x}\mathcal{M})^{s}\rightarrow
\mathbb{R}$  and \emph{a smooth tensorfield of type} $(r,s)$
\emph{on} $\mathcal{M}$ is an $F(\mathcal{M})-$ multilinear map
$(X(\mathcal{M})^{\ast})^{r} \times(X(\mathcal{M}))^{s}\rightarrow
F(\mathcal{M}).$

Thus a $(0,2)$ tensorfield can be identified with a (symmetric)
bilinear form $g(X,Y)$ on vectorfields of $\mathcal{M}.$ This is
called \emph{nondegenerate} provided $g(X,Y)=0$ for all $Y\in
X(\mathcal{M})$ implies $X=0.$ A symmetric bilinear nondegenerate
form $g$ is called \emph{a scalar product.}

At any $x\in\mathcal{M}$ let $e_{1},...,e_{4}$ be a basis of the
tangent space at $x.$ The $4\times4$ matrix
$g_{ab}=g(e_{a},e_{b})$ is called \emph{the matrix of the
tensorfield }$g$ \emph{at} $x$ \emph{relative to the basis}
$e_{1},...,e_{4}.$ Then $g$ is called \emph{Lorentzian} if for
every $x\in\mathcal{M}$ there is a basis of the tangent space
$T_{x}\mathcal{M}$  relative to which the matrix of $g$ at $x$ has
the form $g_{ab}=\textrm{diag}(1,-1,-1,-1).$

It is standard to write the scalar product of two vectorfields
$X^{a}$, $Y^{b}$ using the matrix $g_{ab}$ in the form
$g(X,Y)=g_{ab}X^{a}Y^{b}$. Since $g$ is nondegenerate, $g_{ab}$ is
invertible and we denote the inverse matrix by $g^{ab}.$ Unless
otherwise stated, we use the standard basis of vectorfields
$\partial_{a},$ denote by $X^{a}$ the components of the
vectorfield $X$ with respect to that basis and call $g_{ab}$ the
\emph{components} of the \emph{metric tensor} $g$ relative to the
dual basis of one forms $dx^{a}$ of $\partial_{a},$ that is
$g=g_{ab}dx^{a}dx^{b}.$

Sometimes it proves easier to deal with the function $q$ defined
pointwise on each tangent space by
$q(X)=g(X,X)=g(X^{a}e_{a},X^{b}e_{b})=g_{a b}X^{a}X^{b}$ called
\emph{the associated quadratic form of} $g$. Since the scalar
product $g$ is indefinite there may exist vectorfields $X\neq0$
with $q(X)=0.$\footnote{Such vectorfields are called \emph{null}
and exist only with indefinite scalar products. Two vectorfields
$X$ and $Y$ are called \emph{orthogonal} if $g(X,Y)=0.$ Obviously
a null vectorfield is orthogonal to itself.}
\begin{definition}[Spacetime]
A \textbf{spacetime} is a pair $(\mathcal{M},g)$ where
$\mathcal{M}$ is a manifold and $g$ is a $(0,2)$ tensorfield such
that $\mathcal{M}$ is: 4--dimensional, Hausdorff, connected,
time-oriented and $C^{\infty}$; $g$ is globally defined,
$C^{\infty}$,  nondegenerate and Lorentzian.
\end{definition}
Time-orientability will be defined shortly. In a slight abuse of
notation we often use $\mathcal{M}$ to designate a spacetime. The
simplest example of a spacetime is  the \emph{Minkowski space}
$(\mathbb{R}^{4},\eta_{ab})$  with
$\eta_{ab}=\textrm{diag}(1,-1,-1,-1).$ We note that,
\begin{proposition}
A spacetime is a paracompact manifold.
\end{proposition}
The geometric significance of the Lorentzian metric tensor in a
spacetime $\mathcal{M}$ is reflected in the nontrivial structure
of the tangent spaces of $\mathcal{M}$ at each point and derives
from the following trichotomy.
\begin{definition}[Vector character]
Let $\mathcal{M}$ be a spacetime and $x\in\mathcal{M}.$  We call a
tangent vector $v\in T_{x}\mathcal{M}$
\begin{description}
\item \textbf{spacelike} if $g_{ab}v^{a}v^{b}<0$ (or,
$g(v,v)<0$) or $v=0$
\item \textbf{null} if $g_{ab}v^{a}v^{b}=0$ (or,
$g(v,v)=0$) and $v\neq0$
\item \textbf{timelike} if $g_{ab}v^{a}v^{b}>0$ (or,
$g(v,v)>0$).
\end{description}
We call $v$ a \textbf{causal} vector if $g_{ab}v^{a}v^{b}\geq 0$
(or, $g(v,v)\geq 0$).
\end{definition}
The \textbf{nullcone} at $x$ is the set of all null vectors of
$T_{x}\mathcal{M}.$ The fact that the null cone is in fact a cone
in the tangent space follows easily from the definition. From the
definitions above also follows that timelike vectors are inside
the null cone and spacelike ones are outside. The category into
which a given tangent vector falls is called its \emph{causal
character}. The causal character of any vector $v$ is the same as
the causal character of the subspace $\mathbb{R}v$ it generates. A
subspace $\mathcal{W}$ of $T_{x}\mathcal{M}$ is called
\emph{nondegenerate} if the restriction $g|_{\mathcal{W}}$ is
nondegenerate. A necessary and sufficient condition for the
subspace $\mathcal{W}$ of $T_{x}\mathcal{M}$ to be nondegenerate
is that $T_{x}\mathcal{M}=\mathcal{W}\oplus\mathcal{W}^{\perp}$
where $\mathcal{W}^{\perp} =\{v\in
T_{x}\mathcal{M}:v\perp\mathcal{W}\},$ that is the tangent space
at $x$ is the direct sum of these two subspaces.

In our case (\emph{Lorentz} scalar product) there will always be
degenerate subspaces of $T_{x}\mathcal{M}$, for example, the
subspace spanned by a null vector. In fact there are three
exclusive possibilities for $\mathcal{W}:$
\begin{description}
\item $g|_{\mathcal{W}}$ is positive definite; Then $\mathcal{W}$
is an inner product space and $\mathcal{W}$ in this case is said
to be \emph{spacelike.}
\item $g|_{\mathcal{W}}$ is nondegenerate; Then $\mathcal{W}$ is \emph{timelike}.
\item $g|_{\mathcal{W}}$ is degenerate; Then $\mathcal{W}$ is \emph{null}.
\end{description}
The following lemma is used in the next proposition in an
essential way.
\begin{lemma}
If $u$ is timelike then the subspace $u^{\perp}$ is spacelike.
\end{lemma}
\begin{proof}
From the definition above it follows that since $\mathbb{R}u$ is
timelike $g|_{\mathbb{R}u}$ is nondegenerate and so $\mathbb{R}u$
is nondegenerate. Hence $T_{x}\mathcal{M}=\mathbb{R}u\oplus
u^{\perp}$ and $u^{\perp}$ is nondegenerate. Therefore the index
$\textrm{ind}T_{x}\mathcal{M}=\textrm{ind}\mathbb{R}u+\textrm{ind}u^{\perp}$
i.e., $\textrm{ind}u^{\perp}=0$ that is $u^{\perp}$ is spacelike.
\end{proof}
\begin{proposition}
The null cone disconnects the timelike vectors into two separate
components.
\end{proposition}
\begin{proof}
Let $\mathcal{T}$ be the set of all timelike vectors in
$T_{x}\mathcal{M}.$ For any $u\in\mathcal{T}$ define the
\emph{future timecone} of $u$,
\begin{equation}
\mathcal{C}^{+}(u)=\{v\in\mathcal{T}:g(u,v)>0\},
\end{equation}
and the \emph{past timecone} of $u$,
\begin{equation}
\mathcal{C}^{-}(u)=\{v\in\mathcal{T}:g(u,v)<0\}.
\end{equation}
(Since $u$ is timelike $u\in\mathcal{C}^{+}(u)$ and
$-u\in\mathcal{C}^{-}(u).$) Also obviously
$\mathcal{C}^{+}(u)\cap\mathcal{C}^{-}(u)=\varnothing$ and
$\mathcal{T\subset C}^{+}(u)\cup\mathcal{C}^{-}(u)$ since, for
instance, if $v$ is timelike and does not belong to
$\mathcal{C}^{+}(u)$ then $g(u,v)<0$ and so it belongs to
$\mathcal{C}^{-}(u).$ Finally the case where $v$ is timelike and
orthogonal to $u$, $g(u,v)=0$, is impossible since $u^{\perp}$ is
spacelike according to the preceding lemma.
\end{proof}

Therefore in each  tangent space $T_{x}\mathcal{M}$ of the
spacetime there are two timecones $\mathcal{C}^{+}$ and
$\mathcal{C}^{-}$ which we can arbitrarily call the future
timecone and the past timecone respectively. We call a vector
$u\in\mathcal{C}^{+}$ \emph{future-pointing} and a
$v\in\mathcal{C}^{-}$ \emph{past-pointing}. There is no intrinsic
way to distinguish a future timecone from a past timecone and to
choose one of them is to \emph{time-orient} $T_{x}\mathcal{M}.$

The existence of timecones discussed above raises a fundamental
global question about a spacetime $\mathcal{M}$: \emph{Can every
tangent space in a spacetime be time-oriented in a suitable
continuous way?}
\begin{definition}[Time orientability]
A spacetime $\mathcal{M}$ (or the Lorentz metric $g$ of
$\mathcal{M}$) is called \textbf{time-orientable} if it is
possible to make a consistent continuous choice over all
$\mathcal{M}$ of one set of timelike vectors (say the
future-pointing) at each point of  $\mathcal{M}$.
\end{definition}
If such a choice has been made the spacetime is called
\emph{time-oriented}. In a time-oriented spacetime the null
vectors at each point are also called future-pointing or
past-pointing according as they are the limits of future-pointing
or past-pointing timelike vectors.
\begin{definition}[Time function]
\textbf{A (cosmic) timefunction} on $\mathcal{M}$ is a map
$t:\mathcal{M}\rightarrow
T\mathcal{M}:x\longmapsto\mathcal{C}_{x}$ for each
$x\in\mathcal{M}.$ Such a function maps each point of
$\mathcal{M}$ to a timecone at that point. A timefunction $t$ is
\emph{smooth} if  for every  $x\in\mathcal{M}$ there exists a a
neighborhood $\mathcal{U}$ of $x$ and a vectorfield $X$ on
$\mathcal{U}$ such that for each $y\in\mathcal{U}$ we have
$X_{y}\in\mathcal{C}_{y}.$
\end{definition}
A smooth timefunction on $\mathcal{M}$ is called a
\emph{time-orientation on} $\mathcal{M}$. To choose a specific
time-orientation on $\mathcal{M}$ is to \emph{time-orient}
$\mathcal{M}$.  Thus a spacetime is time-orientable if it admits a
smooth timefunction (i.e., a time-orientation). Once we have
chosen such a function we have time-oriented the spacetime. A
basic criterion for time-orientability of a spacetime is provided
by the following theorem.
\begin{theorem}
Let $\mathcal{M}$ be a spacetime. Then the following two
conditions are equivalent:
\begin{enumerate}
\item $\mathcal{M}$ is time-orientable
\item  There exists a nowhere vanishing, smooth, timelike vectorfield $X$ on
$\mathcal{M}$.
\end{enumerate}
\end{theorem}
Directly related to the notion of time-orientability are the
following two results which show respectively that not every
smooth manifold can be made a spacetime but all spacetimes can
essentially be regarded as time-oriented Lorentz manifolds.
\begin{theorem}
For a smooth manifold $\mathcal{M}$  the following are equivalent:
\begin{enumerate}
\item  There exists a Lorentz metric on $\mathcal{M}$
\item  There exists a time-orientable Lorentz metric on $\mathcal{M}$
\item  There exists a nowhere vanishing vectorfield on $\mathcal{M}$
\item  Either $\mathcal{M}$ is noncompact, or $\mathcal{M}$ is compact and its
Euler characteristic $\chi(\mathcal{M})$ is zero.
\end{enumerate}
\end{theorem}
\begin{theorem}
If a spacetime $\mathcal{M}$ is not time-orientable, there always
exists a double-covering spacetime $\widetilde{\mathcal{M}}$ which
is time-orientable.
\end{theorem}

We now consider paths and curves. A \emph{path} is a continuous
map $\mu:I\rightarrow\mathcal{M}$ where $I$ is an open interval of
$\mathbb{R}.$ A \emph{smooth path} is a path $\mu$ that is smooth
and the differential $d\mu$ is nonzero for all values of the path
parameter which we denote by $t\in I$. A (smooth) \emph{curve}
$\gamma$ is the image of a (smooth) path.
\begin{definition}[Path character]
A smooth path is  called \textbf{timelike (null)} if its tangent
vector at every point is timelike (null). Such a path is called
\textbf{future-oriented} if its tangent vector is future pointing
at every point of the path.  We use the words \textbf{causal path}
for a timelike or a null path.  A \textbf{timelike curve} is the
image of a timelike path. For a timelike curve we write
$\gamma\subset\mathcal{M}.$ Similarly, we speak of a
\textbf{future-oriented, smooth, causal curve}.
\end{definition}
Notice that an arbitrary curve need not have a fixed causal
character.

Next we define the notion of an \emph{endpoint} of a curve. A
\emph{(smooth) curve segment} is a map $\gamma:[a,b]\rightarrow
\mathcal{M}$, where $a=\inf I$ and $b=\sup I$ ($a,b$ can be
$\mp\infty$  respectively), that has a continuous (smooth)
extension to an open interval $J$ containing $[a,b].$ A point $x$
is an endpoint of a curve $\mu$ if  $\mu$ enters and remains in
every neighborhood of $x$. Notice that $x$ is not necessarily a
point on $\mu$. The precise definition is as follows.
\begin{definition}[Endpoints]
 An \textbf{endpoint} of a path or a curve is a point
$x\in\mathcal{M}$ such that, for all sequences $(a_{i})\in I$ such
that $a_{i}\rightarrow a$ we have that $\mu(a_{i})\rightarrow x$,
or if  $a_{i}\rightarrow b$ we have $\mu(a_{i})\rightarrow x.$ If
$\mu$ is causal and future-oriented then the first case defines a
\textbf{past endpoint} whereas the second a \textbf{future
endpoint}. Obviously a causal curve segment is causal at its
endpoints. A timelike curve (or path) without a future (resp.
past) endpoint must extend indefinitely into the future (resp.
past) and is called \textbf{future (resp. past) endless or
inextendible}. A curve that is both future and past endless is
called simply \textbf{endless}.
\end{definition}
An important function associated with any curve in spacetime is
the \emph{length function} which we now define.
\begin{definition}[Length of curves]
Let $\mu$ be a smooth timelike curve in $\mathcal{M}$ with curve
parameter $t\in [0,1]$, endpoints $p=\mu (0)$ and $q=\mu (1)$  and
let $V$ be its tangent vectorfield. We define the \textbf{length
of} $\mu$ from $p$ to $q$ to be the function,
\begin{equation}\label{length}
  L(\mu )=\int_{0}^{1}\bigl( g_{ab}V^{a}V^{b}\bigr) ^{1/2}dt.
\end{equation}
\end{definition}
The length of a null curve is zero and for a spacelike curve we
may take the definition above with a minus sign under the square
root. We return to this function repeatedly below starting with
our discussion of geodesics.

\section{Derivatives}\label{Derivatives}
In this section we discuss collectively various useful derivative
operators defined on the manifold $\mathcal{M}$:
\begin{itemize}
   \item Partial derivative: $\partial_{\mu}$ or $_{,\,\mu}$
   \item Differential of a map $\phi :\mathcal{M}\longmapsto\mathcal{N}$: $\phi_{*}$
   \item Exterior derivative: $d$
   \item Lie derivative: $L_{X}$
   \item Covariant derivative (connection or divergence): $\nabla$ or $_{;\;\mu}$
   \item Covariant derivative along a curve $\gamma (t)$: $\nabla_{t}$ or
   $\frac{D}{\partial t}$
   \item Time derivative: $\dot{V}$
\end{itemize}
For simplicity, we discuss how these operators act on vectorfields
only, leaving their actions of higher rank tensorfields (or
formfields) as an instructive exercise for the reader. We use both
the index-free and index notations for vectors, tensors and forms
invariably according to convenience.

We set as usual,
\begin{equation}\label{pade}
f_{,\;a}\equiv\frac{\partial f}{\partial x^{a}}\;.
\end{equation}
Partial derivatives along vectorfields coincide with the
definition of the action of a vectorfield $X$ of $\mathcal{M}$
with coordinates $X^{a}$ on scalar fields (i.e., functions
$f:\mathcal{M}\rightarrow\mathbb{R}$):
\begin{equation}\label{padealong}
X(f)=X^{a}\partial_{a}f\equiv X^{a}f_{,\;a}\;.
\end{equation}
Consider now a map $\phi :\mathcal{M}\rightarrow\mathcal{N}$ and a
scalar field $f:\mathcal{N}\rightarrow\mathbb{R}$. We define the
\emph{pull back} of the scalar field $f$ to be a function on
$\mathcal{M}$, $\phi^{*}f$, such that,
\begin{equation}\label{pullback}
\phi^{*}f=f\circ\phi\;.
\end{equation}
Hence $\phi^{*}$ pulls back to $\mathcal{M}$ scalar fields defined
on $\mathcal{N}$. For a vectorfield $X$ of $\mathcal{M}$ we define
the derivative of $\phi$, $\phi_{*}$, to be
\begin{equation}\label{derimap}
(\phi_{*}X)(f)=X(\phi^{*}f)\;.
\end{equation}
Thus the derivative of a map between two manifolds maps
vectorfields to vectorfields in the way given above.

The \emph{exterior derivative operator} $d$ maps $r$-formfields
linearly to $(r+1)$-formfields. For example, the coordinate
functions $x^a$ (0-formfields) are mapped to their differentials
$dx^a$ (1-formfields). Consider now an $r$-formfield $\mathbf{A}$
(that is a covariant tensorfield of rank $r$),
\begin{equation}\label{formfield}
\mathbf{A}=A_{\underbrace{{a\dots d}}_{r-times}}\;dx^{a}\dots
dx^{d}\;.
\end{equation}
Then the exterior derivative of $\mathbf{A}$ is the
$(r+1)$-formfield defined by,
\begin{equation}\label{extformfield}
d\mathbf{A}=dA_{\underbrace{{a\dots d}}_{r-times}}\;dx^{a}\dots
dx^{d}\;.
\end{equation}
The chain rule for partial derivatives has an analogue expressed
as the commutation of the exterior derivative and the pull back of
a map, $\phi^{*}$, for an arbitrary $r$-formfield $\mathbf{A}$:
\begin{equation}\label{commutation d *}
d(\phi^{*}\mathbf{A})=\phi^{*}(d\mathbf{A})\;.
\end{equation}

Next consider a \emph{fixed} vectorfield $X$ and its \emph{flow},
$\phi_{t}:\mathcal{M}\longmapsto\mathcal{N},\; t\in\mathbb{R}$,
that is a local 1-parameter group of diffeomorphisms that moves a
point $p\in\mathcal{M}$ a parameter distance $t$ along the
integral curves of $X$ with the properties
$\phi_{t+s}=\phi_{t}+\phi_{s}$, $\phi_{-t}=(\phi_{t})^{-1}$ and
$\phi_{0}=\mathrm{identity}$. Using the derivative $\phi_{t},_{*}$
of the flow we can carry any tensorfield $T$ of $\mathcal{M}$
along the integral curves of the given vectorfield $X$  and
observe how it evolves through the \emph{Lie derivative of the
tensorfield $T$ with respect to the vectorfield $X$} defined by,
\begin{equation}\label{lie}
L_{X}T=\lim_{t\rightarrow 0}\frac{T-\phi_{t},_{*}T}{t}\;.
\end{equation}
(This is minus the Newton quotient.) One may easily show that
$L_{X}$ is a linear map which preserves contractions and tensor
type and satisfies a Leibniz rule. Further it follows from the
definition that on scalar fields we have,
\begin{equation}\label{lie on scalar}
L_{X}f=X(f)\;,
\end{equation}
and also for any vectorfield $Y$ one obtains (exercise),
\begin{equation}\label{poisson like}
(L_{X}Y)f=X(Yf)-Y(Xf)=[X,Y]f=-[Y,X]f\;.
\end{equation}
Thus two vectorfields $X,Y$ \emph{commute} if the Lie derivative
vanishes. In this case if one moves first along $X$ a parameter
distance $t$ \emph{and then} along $Y$ a distance $s$, one arrives
at the same point as if he first goes along $Y$ a distance $s$ and
then along $X$ a parameter distance $t$. This in turn means that
the set of all points so visited forms a 2-dimensional immersed
submanifold through the starting point\footnote{A
$\mathcal{C}^{\infty}$ map  $\phi
:\mathcal{M}\rightarrow\mathcal{N}$ is an immersion if $d\phi$ is
one-to-one. An \emph{imbedding} is an one-to-one immersion.}. From
Eq. (\ref{commutation d *}) it follows that the Lie derivative
also commutes with the exterior derivative for any $r$-formfield
$\mathbf{A}$:
\begin{equation}\label{commutation d lie}
d(L_{X}\mathbf{A})=L_{X}(d\mathbf{A})\;.
\end{equation}
The next derivative operator, the connection, is perhaps the most
important of all. It satisfies:
\begin{theorem}\label{fundamental}
In a spacetime $\mathcal{M}$ there is a unique torsion-free
connection $\nabla$ under which the metric $g$  is covariantly
constant (i.e., parallel).
\end{theorem}
Let $\partial_{a}$ be the standard coordinate vectorfields in the
Minkowski space $\mathbb{M}^{4}.$ As in the case of the Euclidean
space $\mathbb{R}^{n}$, \emph{the covariant derivative}
$\nabla_{V}W$  of a vectorfield $W\in X(\mathbb{M}^{4})$  in the
direction of a fixed $V\in X(\mathbb{M}^{4})$  is given by
\begin{equation}\label{old conn}
  \nabla_{V}W=V(W^{a})\partial_{a}.
\end{equation}
This definition however, does not extend to an arbitrary
spacetime. We define a connection in spacetime by first giving a
new definition of $\nabla_{V}W$ valid in any given manifold
$\mathcal{M}$ and then prove the above theorem in an equivalent
(as we show) form  which has been called \emph{the miracle of
semi-Riemannian geometry.}
\begin{definition}[$\nabla$]
A \textbf{connection} on a manifold $\mathcal{M}$ is a map,
$$\nabla:X(\mathcal{M})\times X(\mathcal{M})\rightarrow
X(\mathcal{M}):(V,W)\longmapsto\nabla_{V}W,$$ such that for all
$V,W,U,S\in X(\mathcal{M}),$ the following properties hold:
\begin{enumerate}
\item [(C1)] $\nabla_{fV+gU}W=f\nabla_{V}W+g\nabla_{U}W$, for all $f,g\in
F(\mathcal{M})$ (hence $\nabla$ is a tensor in the first argument
$V$)
\item[(C2)] $\nabla_{V}(aW+bS)=a\nabla_{V}W+b\nabla_{V}S,$ for
all $a,b\in\mathbb{R}$
\item[(C3)] $\nabla_{V}(fW)=(Vf)W+f\nabla_{V}W,$ for $f\in
F(\mathcal{M}). $
\end{enumerate}
The vectorfield $\nabla_{V}W$ is called the \textbf{covariant
derivative} of $W\in X(\mathcal{M})$ in the direction of $V\in
X(\mathcal{M}).$
\end{definition}
The components of a connection in a coordinate basis have a
special significance.
\begin{definition}[Christoffel symbols]
Let $(x^{a})$ be a coordinate system on $U\subset\mathcal{M}$ and
consider the vectorfield $\nabla_{\partial_{a}}\partial_{b}$ which
gives the rate of change of one coordinate vectorfield relative to
another. Then the components of
$\nabla_{\partial_{a}}\partial_{b}$ in a coordinate basis
$(x^{a})$ are given by,
\begin{equation}
\nabla_{\partial_{a}}\partial_{b}=\Gamma_{ab}^{c}\partial_{c},
\label{christoffel}
\end{equation}
and the functions $\Gamma_{ab}^{c}$ are called \textbf{the
Christoffel symbols in the coordinate system} $(x^{a})$ on $U.$
\end{definition}
A property of the Christoffel symbols which is computationally
advantageous is given in the following proposition.
\begin{proposition}
Let $W=W^{a}\partial_{a}$ be a vectorfield on $\mathcal{M}$ and
$(x^{a})$ a coordinate system on $U\subset\mathcal{M}.$ Then
\begin{equation}
\nabla_{\partial_{a}}(W^{b}\partial_{b})=
\left(\partial_{a}W^{c}+\Gamma_{ab}^{c} W^{b}\right)\partial_{c}.
\label{christ-prop1}
\end{equation}
\end{proposition}
\begin{proof}
Standard application of (C3) with $f=W^{b}.$
\end{proof}
Using (C1) and (\ref{christ-prop1}) we can compute the covariant
derivative $\nabla_{V}W$ for any $V$.

It is not true that on an  arbitrary manifold there exists a
unique connection but in a spacetime the following theorem gives a
uniqueness result for a connection that has three additional
properties.

\begin{theorem}\label{levi-civita}
On a spacetime $\mathcal{M}$ there exists a unique connection
called \textbf{the Levi-Civita connection} $\nabla$ such that for
all $V,W,X\in X(\mathcal{M}),$
\begin{enumerate}
\item [(C4)]$\left[ V,W\right]  =\nabla_{V}W-\nabla_{W}V$\; (torsion-free)

\item[(C5)] $Xg(V,W)=g(\nabla_{X}V,W)+g(V,\nabla_{X}W)$\; (Metric compatible)

\item[(C6)] $2g(\nabla_{V}W,X)=Vg(W,X)+Wg(X,V)-Xg(V,W)\\
-g(V,[W,X])+g(W,[X,V])+g(X,[V,W])$\; (Koszul formula).
\end{enumerate}
\end{theorem}
Condition (C5) is equivalent to the metric $g$ being
\emph{covariantly constant} i.e., $\nabla_{X}g(V,W)=0.$ To see
this, using the product rule we have,
\begin{equation}
\nabla_{X}g(V,W)=(\nabla_{X}g)(V,W)+g(\nabla_{X}V,W)+g(V,\nabla_{X}W),
\end{equation}
and from (C5),
\begin{equation}
\nabla_{X}g(V,W)=(\nabla_{X}g)(V,W)+Xg(V,W) ,\label{*}
\end{equation}
i.e., $(\nabla_{X}g)(V,W)=0$ since on the left-hand side the
\emph{covariant differential }$\nabla_{X}$ is a tensor derivation
operating on the \emph{function} $g(V,W)$, which is just the last
term on the right hand side of  (\ref{*}). Thus this is an
equivalent version of the fundamental theorem \ref{fundamental}.

\begin{proof}[Proof of Theorem \ref{levi-civita}]
Let $\nabla$ be a connection on $\mathcal{M}$ satisfying
properties (C1-C5). Applying (C5) on each of the first three terms
on the right hand side of the Koszul formula, (C4) on each of the
last three terms in the same formula and using the linearity and
symmetry properties of $g$ we find that most terms cancel leaving
the term in the left hand side of (C6). Thus (C1-C5) imply the
Koszul formula for $\nabla$. If now $\overset{\circ}{\nabla}$ is a
second Levi-Civita connection  on $\mathcal{M}$ and we denote by
$F(V,W,X)$ the right hand side of the Koszul formula, then
$2g(\nabla_{V}W,X)=F(V,W,X)$ and therefore
$g(\nabla_{V}W,X)=g(\overset{\circ}{\nabla}_{V}W,X)$ i.e.,
$g(\nabla_{V}W-\overset{\circ}{\nabla}_{V}W,X)=0$ for all $X.$
From the nondegeneracy of the metric $g$ we conclude that the
connection $\nabla$ is unique. To show existence notice that the
function $X\rightarrow F(V,W,X)$ is $F(\mathcal{M})$-linear and so
a one-form. Therefore there exists a unique vectorfield, denote it
by $\nabla_{V}W,$ such that $2g(\nabla_{V}W,X)=F(V,W,X)$ for all
$X.$ Hence the Koszul formula holds for $\nabla _{V}W.$ It is a
computational exercise to show that the Koszul formula implies
properties (C1-C5).
\end{proof}

Let now $\mu:I\rightarrow\mathcal{M}$ be a smooth path on
$\mathcal{M}$ and denote by $X(\mu)$ the set of all vectorfields
along $\mu$. A $V\in X(\mu)$ takes each $t\in I$ to a tangent
vector in $T_{\mu(t)}\mathcal{M}.$
\begin{definition}[$\nabla_{t}$]
The \textbf{covariant derivative of $V$ along $\mu$},
 (also called the induced covariant derivative of
$V$) is a map $$\nabla_{t}:X(\mu)\rightarrow
X(\mu):V\mapsto\nabla_{t}V$$ with the following properties:
\begin{enumerate}
\item [(IC1)]$\nabla_{t}(aV+bW)=a\nabla_{t}V+b\nabla_{t}W,$ for $a,b\in\mathbb{R}$

\item[(IC2)] $\nabla_{t}(fV)=f' V+f\nabla_{t}V,$ for $f\in F(I)$

\item[(IC3)] $(\nabla_{t}W_{\mu})(s)=\nabla_{\mu' (s)}W,$ for all
vectorfields $W\in X(\mathcal{M})$ and $s\in I$.
\end{enumerate}
\end{definition}
This definition is meaningful as we next show.
\begin{proposition}
Let $\mu:I\rightarrow\mathcal{M}$ be a smooth path on
$\mathcal{M}$. Then there exists a unique induced covariant
derivative $\nabla_{t}$ which has the properties (IC1)-(IC3) and
also for every $V,W\in X(\mu)$,
\begin{equation}
\frac{d}{ds}g(V,W)=g(\nabla_{t}V,W)+g(V,\nabla_{t}W). \label{ina}
\end{equation}
\end{proposition}
\begin{proof}
Suppose that an induced covariant derivative $\nabla_{t}$ exists
satisfying (IC1)-(IC3) and assume that the graph of $\mu$ lies in
a single coordinate chart $x^{a}.$ Then for $V\in X(\mu)$ with
coordinates $V^{a}$ we have,
\begin{equation}
V(t)=V^{a}(t)\partial_{a},
\end{equation}
and by (IC1) and (IC2) it follows that on $\mu$,
\begin{equation}
\nabla_{t}V=\frac{dV^{a}}{dt}\partial_{a}+V^{a}\nabla_{t}\partial_{a},
\end{equation}
and since by (IC3) $\nabla_{t}\partial_{a}=\nabla_{\mu
'(s)}\partial_{a}$ we deduce that,
\begin{equation}
\nabla_{t}V=\left(\frac{dV^{a}}{dt}\partial_{a}+V^{a}\nabla_{\mu
'(s)}
\partial_{a}\right)\;, \label{icd}
\end{equation}
which shows that $\nabla_{t}$ is solely determined by the unique
$\nabla.$ Hence uniqueness is proved. Using Eq. (\ref{icd}), it is
not difficult to show that $\nabla_{t}V$ so defined satisfies
properties (IC1)-(IC3) and Eq. (\ref{ina}) and therefore gives a
single vectorfield in $X(\mu)$.
\end{proof}

In the special case for which the vectorfield $V$ along the path
$\mu$ is the tangent vector of $\mu$, $V=\mu'$, the covariant
derivative of $V$ along $\mu$ becomes \emph{the time derivative
of\; $V$}. To see this use (IC3) to write,
\begin{equation}\label{time der cond}
\nabla_{t}V=\nabla_{V}V,
\end{equation}
and calculate the vectorfield $\nabla_{V}V$ in a coordinate basis
to obtain,
\begin{equation}\label{del v v}
\nabla_{V}V=\nabla_{V^{b}\partial_{b}}V^{a}\partial_{a}=V^{b}
\left[ (\partial_{b}V^a
)\partial_{a}+\Gamma^{a}_{cb}V^{c}\partial_{a}\right]
\end{equation}
that is
\begin{equation}\label{del v v final}
\nabla_{V}V=V^{b}\nabla_{b}V^{a}\partial_{a}.
\end{equation}
Using the index notation we thus have,
\begin{equation}\label{time derivative}
\dot{V}^{a}\equiv\nabla_{t}V^{a}=V^b \nabla_{b}V^{a}\;(=V^b
V^{a}_{\;\;\;;b}).
\end{equation}
This has the nice physical interpretation of being \emph{the
acceleration} of a flowline $\mu$ of a fluid and if $\mu$ is taken
to mean the orbit of a particle moving on $\mathcal{M}$,
$\dot{V}^{a}$ denotes the particle's acceleration. The time
derivative of any tensorfield along $\mu$ is  defined similarly
(using the index notation) as,
\begin{equation}\label{time derivative of a tensor}
\dot{T}^{a\dots d}_{e\dots g}=V^h\nabla_{h}T^{a\dots d}_{e\dots
g}\;,
\end{equation}
where $V$ is the tangent vector to the path $\mu$.

\section{Transport and geodesics}
\begin{definition}[Parallel transport]
Let $\mu:I\rightarrow\mathcal{M}$ be a smooth path on
$\mathcal{M}$ and  $V\in X(\mu).$ The vectorfield $V$ is said to
be \textbf{paralelly transported along $\mu$} (or parallel) if,
\begin{equation}
\nabla_{t}V=0.
\end{equation}
\end{definition}
To see this condition expressed in coordinates set $\mu
(t)=x^{a}(t)$ and take $V$ to be the tangent vectorfield,
$V=\dot{x}^{a}$. Then as above $\nabla_{t}$ becomes the time
derivative of $V$ and using Eq. (\ref{del v v final}) we find,
\begin{equation}
\dot{V}^{a}+\Gamma_{bc}^{a}V^{b} V^{c}=0, \label{comp}
\end{equation}
or,
\begin{equation}\label{geodesic equations}
\ddot{x}^{a}+\Gamma_{bc}^{a}\dot{x}^b\dot{x}^{c}=0.
\end{equation}
We see that the condition for parallel transport is equivalent to
a nonlinear system of ODEs -- \emph{the geodesic equations}. By
the fundamental existence and uniqueness theorem valid for such
equations we deduce that for a path $\mu$ and points $p,q\in\mu$
one obtains a unique vector at $q$ by parallelly transporting a
given vector $v$ at $p$ along $\mu$. Here by \emph{parallel
transport along} $\mu$ we mean a map,
\begin{equation}\label{par trans map}
P_{a}^{b}(\mu):T_{\mu(a)}\mathcal{M}\rightarrow
T_{\mu(b)}\mathcal{M}:v\mapsto V(b),
\end{equation}
where the vectorfield $V$ is parallel and $V(a)=v$. It follows
easily that $P_{a}^{b}$ is a linear isomorphism and from
(\ref{ina}), if $V,W$ are parallel, it follows that $g(V,W)$ is
constant. Hence taking two vectors at $p$ ($v,w\in
T_{p}\mathcal{M}$ with $V(a)=v$ and $W(a)=w$) we obtain,
\begin{equation}
g(P_{a}^{b}(v),P_{a}^{b}(w))=g(V(b),W(b))=g(V(a),W(a))=g(v,w).
\label{par-trans}
\end{equation}
Therefore we have shown that the fundamental Theorem
\ref{fundamental} takes still another equivalent form as follows:
\begin{theorem}\label{scalar product preservation}
In a spacetime $\mathcal{M}$ parallel transport along any curve
preserves the scalar product defined by the metric $g$.
\end{theorem}
It is also true that in an arbitrary spacetime despite the
situation in the Minkowski space $\mathbb{M}^{4}$  wherein one is
using the same natural coordinates (distant parallelism), parallel
transport \emph{depends on the particular path} $\mu$ that we are
using to move a vector placed at an initial point  $p$ to a final
point $q$ via a parallel vectorfield. Equivalently, around a
closed curve, the final vector $w$ obtained by parallelly
transporting an initial vector $v$ need not be $v$, a phenomenon
called \emph{holonomy}. In this case all possible parallel
vectorfields along $\mu$ are rotated through an angle called the
holonomy angle.

As discussed above the induced covariant derivative operator
$\nabla_{t}$ can be trivially  applied to the tangent vectorfield
$\mu' (s)$  of a path $\mu$  to give the vectorfield
$\nabla_{t}\mu' $ called \emph{the acceleration of the path
}$\mu$. It is tempting sometimes to write for a vectorfield $V\in
X(\mu)$, $\nabla_{\mu' }V=\nabla_{t}V.$  In this notation,
$\nabla_{t}\mu' =\nabla_{\mu' }\mu' $ which we sometimes
abbreviate to $\mu''.$ We shall see that this operation gives us
important global information about the behaviour of a path.
\begin{definition}[Geodesics]
A \textbf{geodesic} in a spacetime $\mathcal{M}$  is a path
$\mu:I\rightarrow\mathcal{M}$ such that for every $s\in I$,
\begin{equation}
  \nabla_{\mu'}\mu' =0,
\end{equation}
that is $\mu$ has zero acceleration. Equivalently, a geodesic is a
path such that its tangent vectorfield is parallel.
\end{definition}
Thus geodesics satisfy (i.e., are solutions of) the geodesic
equations (\ref{geodesic equations}). Geodesics have quite uniform
behaviour. Every constant $(\mu' =0)$ curve is trivially a
geodesic called a degenerate geodesic, but if for an $s_{0}\in I$,
$\mu' (s_{0})\neq0$ for a geodesic $\mu$ then, since geodesics by
definition have constant speed, $\mu' (s)\neq 0$ for every $s\in
I.$  In this case $\mu$ is called \emph{a nondegenerate geodesic}.
In what follows all geodesics will be assumed nondegenerate. Thus
a geodesic cannot slow down and stop.

Note also that the \emph{affine parameter} $s$ of a geodesic $\mu$
is determined only up to transformations of the  form
$s\rightarrow as+b$ where $a,b$ are constants ($a$ corresponds to
renormalizations of $\mu' $ of the form $\mu' \rightarrow(1/a)\mu'
$ and $b$ to the freedom of choosing the initial point $\mu(0)).$
All these affine parameters define the same geodesic.
\begin{definition}[Geodesic character]
A  geodesic $\mu$ is called \textbf{timelike,  null, spacelike or
causal} if $\mu' $ is timelike,  null, spacelike or causal
respectively.
\end{definition}
Notice that unlike an arbitrary  curve a geodesic $\mu$ has
necessarily one of the three causal characters. For if at an
$s_{0}$ the tangent vector $\mu'$ is timelike ($g(\mu',\mu')>0)$
then, since by Theorem \ref{scalar product preservation} parallel
transport preserves the scalar product defined by the metric $g$,
$\mu'$ will always stay timelike.

Using the common abbreviation $x^{a}\circ\mu=x^{a}$ it follows
from the existence and uniqueness theorem that the geodesic
equations,
\begin{equation}
\ddot{x}^{a}+\Gamma_{bc}^{a}\dot{x}^b\dot{x}^{c}=0,
\end{equation}
yield for any $v\in T_{x}\mathcal{M}$ a unique  geodesic
$\mu_{v}:I_{v}\rightarrow\mathcal{M}$ that passes through $x$
i.e., $\mu _{v}(0)=x $ and has initial velocity $\mu_{v}' (0)=v$.
We can therefore talk of \emph{a geodesic starting at} $x$ with
initial velocity $v$. We also deduce that the domain $I_{v}$ is
the largest possible. Because  of this $\mu_{v}$ is called
\emph{maximal} or \emph{inextendible.} We therefore arrive at the
following definition.
\begin{definition}[g-completeness]
A spacetime $\mathcal{M}$ is called \textbf{geodesically
complete}, if every maximal geodesic is defined on the entire real
line. In this case we speak of \textbf{complete geodesics}.
\end{definition}
For a different characterization and further properties of
complete spacetimes we need to introduce some simple properties of
the exponential map. Let $x\in T_{x}\mathcal{M}$  and consider the
subset $\Delta_{x}$ of $T_{x}\mathcal{M}$ consisting of those
$v\in T_{x}\mathcal{M}$ such that the unique inextendible geodesic
$\mu_{v}$ is defined at least on $[0,1]$ (that is $I_{v}\supset
[0,1]$).
\begin{definition}[Exp map]
By the \textbf{exponential map} of $\mathcal{M}$ at $x$ we mean a
map,
\begin{equation}
\exp_{x}:\Delta_{x}\rightarrow\mathcal{M}:v\longmapsto\exp_{x}
(v)=\mu_{v}(1).
\end{equation}
\end{definition}
If for every $x\in\mathcal{M},$ $\Delta_{x}= T_{x}\mathcal{M}$,
then every inextendible geodesic of $M$ is defined on the whole
real line that is $\mathcal{M}$ is geodesically complete and so
every inextendible geodesic extends to arbitrary parameter values.

The geometric meaning of the exponential map is obtained from the
following result.
\begin{lemma}
For any $x\in\mathcal{M}$ the map $\exp_{x}$ carries radial lines
through the origin of $T_{x}\mathcal{M}$ to geodesics through $x$,
that is $\exp _{x}(tv)=\mu_{v}(t).$
\end{lemma}
\begin{proof}
Consider a fixed $t\in \mathbb{R}$ and a fixed $v\in
T_{x}\mathcal{M}$. Then the geodesic path
$s\longmapsto\mu_{v}(ts)$ has initial velocity $t\mu_{v}' (0)$,
that is $tv$ which is obviously the initial velocity of
$\mu_{tv}(s).$ Thus for all $s$ and $t$ we have
$\mu_{v}(ts)=\mu_{tv}(s)$, and setting $s=1$ we find,
\begin{equation}
\exp_{x}(tv)=\mu_{tv}(1)=\mu_{v}(t),
\end{equation}
and this completes the proof of the Lemma.
\end{proof}

It follows  that the exponential map at a point on a spacetime
collects together all geodesics passing through that point. Also
if $\mathcal{Q}$ is any neighborhood of the origin of
$T_{x}\mathcal{M}$, then $\exp(\mathcal{Q})$ is a neighborhood of
$x$ diffeomorphic to $\mathcal{Q}$ as the following Lemma shows.
\begin{lemma}
For every $x\in\mathcal{M}$ the exponential map at $x$ carries a
neighborhood $\mathcal{Q}$ of the origin $0\in T_{x}\mathcal{M}$
diffeomorphically onto a neighborhood $\mathcal{N}$ of $x$ in
$\mathcal{M}.$
\end{lemma}
\begin{proof}
This is easy using the inverse function theorem. It suffices to
show that the differential map,
\begin{equation}
d\exp_{x}:T_{0}(T_{x}\mathcal{M})\rightarrow T_{x}\mathcal{M},
\end{equation}
is the linear isomorphism $v_{0}\mapsto v$. If we set $\rho(t)=tv$
and $v_{0}=\rho'(0)$ the claim follows from the fact that
\begin{equation}
d\exp_{x}(v_{0})=d\exp_{x}(\rho' (0))=(\exp_{x}\circ\rho)^{'
}(0)=\mu_{v}' (0)=v.
\end{equation}
The proof is now complete.
\end{proof}

In the following we assume that $\mathcal{Q}$ is \emph{starshaped}
about $0$ in $T_{x}\mathcal{M}$, that is for each $\lambda\in
[0,1]$ the segments $\lambda v\in\mathcal{Q}.$ In this case the
neighborhood $\mathcal{N}$ in the above lemma is called a
\emph{normal neighborhood of }$x$. When $\mathcal{Q}$ is an open
disk of radius $\varepsilon$ we speak of a normal neighborhood of
radius $\varepsilon$ and write $\mathcal{N}_{\varepsilon}.$ We now
give without proof the characteristic property of normal
neighborhoods.
\begin{proposition}\label{normal nei}
If $\mathcal{N}$ is a normal neighborhood of $x\in\mathcal{M}$,
then for each $p\in\mathcal{N}$ there exists a unique geodesic
path $\mu:[0,1]\rightarrow\mathcal{N}$ joining $x$ and $p$ with
initial velocity $\mu' (0)=\exp^{-1}(x)$ in $\mathcal{Q}$. It
follows that $\mathcal{N}$ uniquely describes $\mathcal{Q}$.
\end{proposition}
The part of a spacetime in a normal neighborhood of a point can be
described in such a way that the length of a geodesic emanating
from the point to one inside the normal neighborhood is given in a
very simple form. To show this let $e_{(a)}$ be a \emph{frame}
(orthonormal basis) of $T_{x}\mathcal{M}$ and $\mathcal{N}$ a
normal neighborhood of $x.$ The \emph{normal coordinates} $x^{a}$
\emph{of a point} $p\in\mathcal{N}$ are the coordinates of the
tangent vector $\exp_{x}^{-1}(p)\in\mathcal{Q}\;(\subset
T_{x}\mathcal{M)}$ relative to the frame $e_{(a)}$, that is
$\exp_{x}^{-1}(p)=x^{a}$ $e_{(a)}.$\footnote{At $x$ the metric
$g_{ab}(x)$ in terms of the normal coordinates takes the form
$g_{ab}=\textrm{diag}(1,-1,-1,-1)$ and thus
$\Gamma_{ab}^{\rho}(x)=0$. This also implies  that covariant
differentiation of any tensor at $x$ is replaced, when normal
coordinates at $x$ are used, by common partial differentiation in
terms of these normal coordinates.}

Suppose now that a geodesic emanating from $x$ to $p$ has initial
velocity $\mu '=v.$  The arc length of $\mu$ is
$L(\mu)=\int_{0}^{1}|\mu '(s)| ds$, where $\mu '(s)| ^{2}
=g_{ab}(x^{a}\circ\mu )'(x^{b}\circ\mu)'.$ We define \emph{the
radius function of} $\mathcal{M}$ \emph{at} $x$,
\begin{equation}
r(p)=\left|\exp_{x}^{-1}(p)\right| ,
\end{equation}
which in normal coordinates is just
$\left((x^{1})^{2}-(g_{ab}x^{a}x^{b})^{2}\right)^{1/2}.$ Since the
exponential map sends lines to geodesics this definition uses
implicitly the fact that the tangent vector to the geodesic $\mu$
at $p$ is $\exp_{x}^{-1}(p).$ The radius function is thus smooth
at all points except at $x$. Since $\mu '(s)$ is constant (arc
length parametrization) we conclude that in these coordinates
$L(\mu)=r(p)=|v|.$ The radius function is positive, zero or
negative according to whether the geodesic $\mu$ from $x$ to $p$
is timelike, null or spacelike (that is $\exp_{x}^{-1}(p)$ is!)
respectively.
\begin{definition}[Simply convex]
Let $\mathcal{M}$ be a spacetime. A set $\mathcal{N\subset M}$ is
called \textbf{simply convex} if it is a normal neighborhood of
each of its points. $\mathcal{N}$ is called a \textbf{simple
region} if it is open, simply convex and the closure
$\overline{\mathcal{N}}$ is a compact set contained in a simply
convex, open set in $\mathcal{M}$.
\end{definition}
The entire manifold of Minkowski spacetime is simply convex. The
characteristic property of a simply convex neighborhood
$\mathcal{N}$ is that there exists a unique geodesic lying
entirely in $\mathcal{N}$ connecting any pair of points
$p,q\in\mathcal{N}$ as it follows by applying Proposition
\ref{normal nei}. This nice local behaviour of geodesics in a
normal neighborhood is in sharp contrast to what may happen
globally in an arbitrary spacetime. In general there may exist
points that can be connected by no geodesic or geodesics passing
through a point may focus at some other point in a spacetime.

\section{Conjugate points and geodesic congruences}\label{conjugate
points and geodesic congruences} The Riemann curvature tensor in
this paper is determined by the equation,
\begin{equation}
R(V,U)W=\nabla_{[V,U]}W-[\nabla_{V},\nabla_{U}]W,
\end{equation}
for any three vectorfields $V,U,W\in X(\mathcal{M}).$ In this
Section we discuss the problem of how to compare nearby geodesics.
We start by introducing the notion of \emph{a 1-parameter family
of geodesics}. This is described by a map,
\begin{equation}
x:[a,b]\times(-\varepsilon,\varepsilon)\rightarrow\mathcal{M}:(t,u)
\longmapsto x(t,u).
\end{equation}
We understand that for each constant value of the parameter $u$ we
have a geodesic parametrized by $t$. Each such \emph{longitudinal}
geodesic has velocity $x_{t}$ and acceleration $x_{tt}=0.$ The
\emph{transverse} curves $u\mapsto x(t,u)$ have velocity $x_{u}$.

Consider now the \emph{variation vectorfield} (or the vectorfield
of geodesic variation) $x_{u}(t,0)\equiv (d/du)|_{u=0}x_{u}(t,u)$
on the longitudinal geodesic $x(t,0)$ (called the base or initial
geodesic) which is a vectorfield along the geodesic. Since
$x_{tu}=x_{ut}$ we have,
\begin{equation}
x_{utt}=x_{tut}=x_{ttu} +R(x_{u},x_{t})x_{t}=R(x_{u},x_{t})x_{t},
\end{equation}
i.e., the vectorfield $x_{u}$, call it $Y$, satisfies the linear
second-order equation,
\begin{equation}
x_{utt}=R(x_{u},x_{t})x_{t}.
\end{equation}
This motivates the following more general notion.
\begin{definition}[Jacobi field]
Let $\mu$ be a geodesic and $Y$ a vectorfield along $\mu.$ We say
that $Y$ is a \textbf{Jacobi vectorfield on} $\mu$ if $Y$
satisfies the \textbf{Jacobi differential equation},
\begin{equation}\label{jacobi eqn}
\nabla_{tt}Y=R(Y,\mu' )\mu' .
\end{equation}
\end{definition}
This definition and the calculation above show that the variation
vectorfield of the base geodesic in the $1$-parameter family  of
geodesics is a Jacobi field. A way of interpreting the Jacobi
equation that appears often in applications is to show that for a
given vectorfield $Y$ the values of both sides of the Jacobi
equation are the same. Intuitively a Jacobi field connects points
of the geodesic $\mu$ with corresponding points of a neighboring
geodesic $\nu.$

The Jacobi equation is a linear differential equation and
therefore the set of all Jacobi fields forms a real linear space.
The dimension of this space is twice the dimension of
$T_{x}\mathcal{M}$ since any solution of the Jacobi equation is
defined by specifying (arbitrarily) the value of the vector $Y$
and that of the vector $\nabla_{t}Y$ at any point on the geodesic.
\begin{definition}[Conjugate points]
We say that two points $p=\mu(a)$ and $q=\mu(b)$ on a geodesic
$\mu$ are \textbf{conjugate along} $\mu$ provided there is a
nonzero Jacobi field $J$ on $\mu$ such that $J(a)=0$ and $J(b)=0.$
\end{definition}
Roughly speaking, a pair of conjugate points occurs when two
neighboring geodesics meet at $p$ and then refocus at
$q$\footnote{It is easily seen that the normal coordinate system
defined by $\exp_{p}$ becomes singular at $q.$}. We can obtain a
more general notion of conjugation if we replace one point in a
pair of conjugate points  by a \emph{submanifold} of our
spacetime. We restrict attention to the case of a \emph{spacelike
three-surface} $\Sigma$ imbedded in the spacetime $\mathcal{M}$.
Thus $\Sigma$ may be thought of as the three-dimensional graph of
a $\mathcal{C}^{2}$ function $f$ defined locally by $f=0$ and such
that $g^{ab}\nabla_{a}f\nabla_{b}f>0$ when $f=0$. Consider then a
congruence (see below) of timelike geodesics \emph{meeting
$\Sigma$ orthogonally}. In this case we have the following
definition.
\begin{definition}[Focal points]
A point $p$ on a geodesic $\gamma$ of a geodesic congruence
orthogonal to $\Sigma$ is called \textbf{conjugate to $\Sigma$
along $\gamma$} (or a focal point) if there exists a  Jacobi field
along $\gamma$ which is nonzero on $\Sigma$ but vanishes at $p$.
\end{definition}
We shall see below that geodesics are length-maximizing curves in
a spacetime when they do not possess conjugate points. It is
therefore basic to understand the precise conditions under which a
pair of conjugate points will exist in a spacetime $\mathcal{M}$.
Consider for this purpose a \emph{congruence} of curves on
spacetime. A congruence is a bunch of curves such that through
each point of spacetime there passes precisely one curve from this
bunch. A physically plausible way of visualizing this is to think
of a congruence of curves as the flowlines of a fluid flow.  Given
a congruence of curves the tangent field is a well-defined
vectorfield and conversely, it can be shown that every continuous
vectorfield of $\mathcal{M}$ generates a congruence of curves. In
the following we focus our attention to congruences of timelike
geodesics. The theory of congruences of null geodesics, although
analogous, is different and will not be treated here.

Consider a congruence of \emph{timelike} geodesics and the
cross-sectional area thought of as obtained by cutting the
flowlines by a plane and taking the area enclosed by a small
circle around the bunch. From the Jacobi equation through an
argument which we omit it follows that, due to the tendency of
geodesics to accelerate towards or away from each other, as the
geodesic flowlines `move' in spacetime this area can do three
things: It can \emph{expand} (or contract), it can \emph{get
distorted or sheared}, so that the circle enclosing it becomes an
ellipse, or it can \emph{twist} (i.e., rotate). How do we describe
these three possible changes?

In the singularity theorems-related literature, the most common
method to proceed is to form the covariant derivative
$\nabla_{b}V_{a}$ of the tangent vectorfield $V_{a}\;
(=g_{ab}V^{b})$ to the flowlines of the geodesic congruence and
eventually obtain a set of equations describing the evolution of
the expansion and the other parameters defined above (the
Raychaudhuri equations-see below). $\nabla_{b}V_{a}$  is the
gradient of the fluid velocity vectorfield, and the aforementioned
changes in the cross-sectional area during the evolution of the
congruence are reflected in this derivative, for \emph{this
measures the failure of the displacement vector between two
neighboring geodesics to be parallelly transported}. To see this
consider again the steps in the derivation of the Jacobi equation
(\ref{jacobi eqn}) and notice that the two coordinate vectorfields
$\partial /\partial t$ and $\partial /\partial u$ (the second
giving the displacement vector at each point) commute, that is if
we call $V=\partial /\partial t$ and $Z=\partial /\partial u$ we
find,
\begin{equation}\label{failure}
  V^{b}\nabla_{b}Z^{a}=  Z^{b}\nabla_{b}V^{a},
\end{equation}
which means that the failure of the RHS to be zero is equivalent
to the failure of the connecting vector $Z$ is be parallelly
transported (i.e., LHS zero) along the geodesic flow. An intuitive
way to say this is that an observer on the base geodesic will see
nearby geodesics to be stretched and twisted.

Let $V$ be the unit tangent timelike vectorfield to the congruence
($g(V,V)=1$) as above. This vectorfield clearly defines, at each
point $q$ of the flow lines, a space $H_{q}$ orthogonal to it
($V_{q}\in T_{q}\mathcal{M}\;\textrm{and}\; H_{q}\subset
T_{q}\mathcal{M})$. Therefore we can take any vector $X_{q}\in
T_{q}\mathcal{M}$ and project it in the direction orthogonal to
$V_{q}$, that is in $H_{q}$. This can be done through \emph{the
projection operator} $h^{a}\,_{b}=g^{ac}h_{cb}$ where,
\begin{equation}\label{proj oper}
  h_{ab}=g_{ab}+V_{a}V_{b}.
\end{equation}
It can be shown that,
\begin{equation}\label{splitting}
\nabla_{b}V_{a}=\frac{1}{3}\theta h_{ab}+\sigma_{ab}+\omega_{ab},
\end{equation}
where we define the \emph{expansion} $\theta$, the \emph{shear}
$\sigma_{ab}$ and the \emph{twist} $\omega_{ab}$ of the geodesic
congruence by
\begin{eqnarray}
\theta&=&h^{ab}\nabla_{b}V_{a},\label{exp}\\
\sigma_{ab}&=&\nabla_{(b}V_{a)}-\frac{1}{3}\theta
h_{ab},\label{shear}\\ \omega_{ab}&=&\nabla_{[
b}V_{a]}\label{twist}.
\end{eqnarray}
To proceed further we understand that the basic quantities
$\theta$, $\sigma_{ab}$ and $\omega_{ab}$ are functions of time
and the question arises as to what are the equations that describe
the evolution of these quantities. The resulting differential
equations are the \emph{Raychaudhuri equations} and play a central
role in the proofs of the singularity theorems. We derive only the
first such equation namely, the one that gives the evolution of
the expansion $\theta$ as this is the most important for our
purposes.

From our definition of the Riemann curvature tensor it follows
that for any vectorfield $V$,
\begin{equation}\label{riemann}
  \bigl(\nabla_{b}\nabla_{c}-\nabla_{c}\nabla_{b}\bigr)
  V_{a}=-R_{bca}^{\;\;\;\;d}V_{d},
\end{equation}
and so for our timelike $V$ we can  calculate the time derivative
of the tensorfield $\nabla_{b}V_{a}$,
\begin{eqnarray}\label{t deriv of the basic tensor}
V^{c}\nabla_{c}(\nabla_{b}V_{a})&=&V^{c}\nabla_{b}(\nabla_{c}V_{a})+
R_{bca}^{\;\;\;\;d}V^{c}V_{d}\\ \nonumber
&=&\nabla_{b}(V^{c}\nabla_{c}V_{a})-(\nabla_{b}V^{c})(\nabla_{c}V_{a})+
R_{bca}^{\;\;\;\;d}V^{c}V_{d}\\ \nonumber
&=&-(\nabla_{b}V^{c})(\nabla_{c}V_{a})+
R_{bca}^{\;\;\;\;d}V^{c}V_{d}.
\end{eqnarray}
since the first term in the second line above contains the time
derivative of the tangent vectorfield to the geodesic and is
therefore zero. Tracing this equation with $h^{ab}$ and using the
definitions (\ref{exp}-\ref{twist}) we find,
\begin{equation}\label{raycha}
V^{c}\,\nabla_{c}\,\theta\equiv\dot{\theta}=-\frac{1}{3}\,\theta^{2}-
\sigma_{ab}\,\sigma^{ab}+\omega_{ab}\,\omega^{ab}-R_{cd}\,V^{c}\,V^{d}.
\end{equation}
This is the \emph{Raychaudhuri equation} for the evolution of the
expansion scalar $\theta$. Defining the \emph{scalar shear}
$\sigma$ and the \emph{scalar twist} $\omega$ through the
equations,
\begin{equation}\label{scalar kin}
  2\sigma^{2} =\sigma_{ab}\,\sigma^{ab}>0,\quad 2\omega^{2}
  =\omega_{ab}\,\omega^{ab}>0,
\end{equation}
we can write the Raychaudhuri equation more compactly as,
\begin{equation}\label{raycha'}
  \dot{\theta}=-\frac{1}{3}\,\theta^{2}-
  2\sigma^{2}+2\omega^{2}-R_{ab}\,V^{a}\,V^{b}.
\end{equation}
Suppose now that the congruence meets $\Sigma$ orthogonally (or
that the curves in the congruence start from a point $p$). In this
case the twist vanishes, $\omega_{ab}=0$ (this follows from
Frobenius theorem - see \cite{wa84}, page 435), and the
Raychaudhuri equation (\ref{raycha'})  becomes,
\begin{equation}\label{omega zero}
  \dot{\theta}=-\frac{1}{3}\,\theta^{2}-
  2\sigma^{2}-R_{ab}\,V^{a}\,V^{b}.
\end{equation}
This shows that \emph{provided} $R_{ab}\,V^{a}\,V^{b}\geq 0$ all
terms in the RHS are negative and so,
\begin{equation}\label{solve1}
\dot{\theta}+\frac{1}{3}\,\theta^{2}\leq 0,\quad\
(\theta^{-1})^{\cdot}\geq \frac{1}{3},
\end{equation}
or
\begin{equation}\label{solve2}
  \frac{1}{\theta}\geq  \frac{1}{\theta_{0}}+\frac{1}{3}t,
\end{equation}
where $\theta_{0}$ is the initial value of $\theta$. Therefore if
we suppose that the congruence is initially converging,
$\theta_{0}<0$, then within a time
\begin{equation}\label{time for conj point}
  t\leq\frac{3}{|\theta_{0}|},
\end{equation}
the expansion  becomes infinite, $\theta\rightarrow -\infty$, that
is there is a second conjugate point, say $q$, to $p$ or a focal
point to $\Sigma$. In other words negative expansion implies
refocusing or convergence of the geodesic congruence. This
motivates the following definition.
\begin{definition}[TCC]\label{cc}
We say that a spacetime satisfies the \textbf{timelike convergence
condition} if
\begin{equation}\label{tcc}
  R_{ab}\,V^{a}\,V^{b}\geq 0,\quad\textrm{for all timelike
  vectorfields}\; V^{a}.
\end{equation}
\end{definition}
If this holds for all null vectorfields $V^{a}$, then we call it
the \emph{null convergence condition}. By continuity the timelike
implies the null convergence condition.

In the case when $R_{ab}\,V^{a}\,V^{b}=0$ everywhere on $\gamma$,
one can show that provided the tidal forces are not zero,
$R_{abcd}\,V^{a}\,V^{d}\neq 0$, the shear term in the Raychaudhuri
equation cannot vanish and therefore a similar argument as above
establishes the existence of point conjugate to a point or a
hypersurface. The above condition on the Riemann tensor is very
important. We frame it in a definition which also includes the
null case.
\begin{definition}[GC]
We say that a spacetime satisfies the \textbf{timelike generic
condition} if
\begin{equation}\label{tgc}
R_{abcd}\,V^{b}\,V^{c}\neq 0.
\end{equation}
We say it satisfies the \textbf{null generic condition} if
\begin{equation}\label{ngc}
V_{[a}R_{b]cd[e}V_{f]}V^{c}\,V^{d}\neq 0 .
\end{equation}
In such cases, we speak of a \textbf{generic vectorfield} $V^{a}$.
\end{definition}
When a vectorfield fails to satisfy a generic condition we call it
\emph{nongeneric}.

In the discussion above we have provided conclusive evidence (but
not the proof which we omit) for the following result.
\begin{theorem}\label{conj point thm}
Let $(\mathcal{M},g)$ be a spacetime in which
$R_{ab}\,V^{a}\,V^{b}\geq 0$ and also $R_{abcd}\,V^{a}\,V^{d}\neq
0$ for all timelike $V$. Then every complete timelike geodesic
possesses a pair of conjugate points. Also if $\Sigma$ is a
spacelike hypersurface with $\theta_{0}<0$ at some point
$q\in\Sigma$, then within a time $t\leq 3/|C|$, $C$ constant,
there exists a point $p$ conjugate to $\Sigma$ along a geodesic
$\gamma$ orthogonal to $\Sigma$ passing through $q$ assuming that
$\gamma$ can be extended to these values.
\end{theorem}
Recall that an important property of geodesics is that they
extremize the length function (\ref{length}) among all possible
curves connecting two points in spacetime. In fact, using the
second variation of the length function we can show that this
extremum is a maximum if and only if there are no conjugate points
between the endpoints of the family.
\begin{theorem}\label{max}
Consider a 1-parameter family of smooth timelike curves connecting
two points $p$ and $q$. Then  the length function $L(\gamma )$ has
a maximum on a curve $\gamma$ iff this curve satisfies the
geodesic equations (\ref{geodesic equations}) with no points
conjugate to $p$ between $p$ and $q$.
\end{theorem}
In analogy with the point to point conjugation, the absence of
conjugate points to the hypersurface $\Sigma$ provides a necessary
and sufficient condition for length maximization.
\begin{theorem}\label{max1}
Consider a 1-parameter family of smooth timelike curves connecting
a point $p$ to $\Sigma$. Then the length function $L(\gamma )$ has
a maximum on a curve $\gamma$ iff this curve satisfies the
geodesic equations (\ref{geodesic equations}) with no points
conjugate to $\Sigma$ between $p$ and $\Sigma$.
\end{theorem}
Before we proceed further we make the last definition of this
Section. Let $x$ be a $1$-parameter family of geodesics through
$p.$ The set of all points of $\mathcal{M}$ conjugate to $p$ along
geodesics from the family $x$ is called a \emph{caustic.} A
caustic is, roughly speaking, the locus of points where
consecutive geodesics intersect.

\section{Causal geometry}\label{causal structure}
We shall use the word \emph{trip} to indicate a curve that is
piecewise a future--pointing, timelike geodesic and we understand
that for any two points $x,y\in\mathcal{M}$ \emph{a trip from $x$
to $y$} is a trip with past endpoint $x$ and future endpoint $y.$
Similarly we define a \emph{causal trip} to mean a curve that is
piecewise a future pointing, causal (timelike or null) geodesic,
possibly degenerate. Then two basic relations can be defined on
$\mathcal{M}.$
\begin{definition}[Causality and chronology]
Let $\mathcal{M}$ be a spacetime and $x,y\in\mathcal{M}.$ We say
that $x$ \textbf{chronologically precedes} $y,$ $x\ll y,$ if and
only if there is a trip from $x$ to $y.$
\begin{equation}\label{chronology}
x\ll y \Leftrightarrow \textrm{there is a trip from}\; x\;
\textrm{to} \;y.
\end{equation}
We call the relation $\ll$ a \textbf{chronology} relation on
$\mathcal{M}$. We say that $x$ \textbf{causally precedes} $y$,
$x\prec y,$ if and only if there is a causal trip from $x$ to $y$.
\begin{equation}\label{causality}
x\prec y\Leftrightarrow \textrm{there is a causal trip from} \;x
\;\textrm{to} \;y.
\end{equation}
We call the relation $\prec$ a \textbf{causality} relation on
$\mathcal{M}$.
\end{definition}
Evidently, $x\ll y$ implies $x\prec y.$ Also, chronology and
causality are transitive relations. Since degenerate causal
geodesics are allowed it follows that $x\prec x$ is possible but
in contrast $x\ll x$ means that there exists a closed trip with
past and future endpoint $x$. A closed, nondegenerate causal trip
connecting two distinct points $x,y$ is signified by $x\prec y$
and $y\prec x.$

The following definition gathers together the points $y$ of
$\mathcal{M}$ that can be influenced by, or influence, a given
point $x\in\mathcal{M}$. It is the fundamental definition of
causal structure theory.
\begin{definition}[Futures and pasts]
Let $x$ be a fixed point in $\mathcal{M}$. The
\textbf{chronological future of} $x$ is the set
\begin{equation}
I^{+}(x)=\Bigl\{ y\in\mathcal{M}:x\ll y\Bigr\} .
\end{equation}
The \textbf{chronological past of }$x$ is the set
\begin{equation}
I^{-}(x)=\Bigl\{  y\in\mathcal{M}:y\ll x\Bigr\} .
\end{equation}
The \textbf{causal future of }$x$ is the set
\begin{equation}
J^{+}(x)=\Bigl\{  y\in\mathcal{M}:x\prec y\Bigr\} .
\end{equation}
The \textbf{causal past of }$x$ is the set
\begin{equation}
J^{-}(x)=\Bigl\{  y\in\mathcal{M}:y\prec x\Bigr\} .
\end{equation}
For any given set $\mathcal{S\subset M}$ we define its
chronological future $I^{+}(\mathcal{S})=\cup_{x\in\mathcal{S}}
I^{+}(x)$ and similarly for the other ones.
\end{definition}
The dual versions of any result are obvious and will always be
assumed. We think of $I^{+}(x)$ as the set of all events in
spacetime that can be influenced by what happens at $x$ and
similarly for the other definitions. We call a spacetime
\emph{chronological (resp. causal)} if for every $x\in\mathcal{M}$
we have $x\notin I^{+}(x)$ (resp. $x\notin J^{+}(x)$). Thus in a
chronological (causal) spacetime there are no closed timelike
(causal) curves. A spacetime is called \emph{distinguishing} if
for any two points $x,y\in\mathcal{M}$ we have $x\neq y$
$\Rightarrow$ $I^{\pm}(x)\neq I^{\pm}(y)$. In a distinguishing
spacetime points are distinguished by their chronological futures
and pasts.

The following Proposition gives some simple properties of futures
and pasts.
\begin{proposition}\label{f-t}
Let $x$ be any point in $\mathcal{M}$ and $\mathcal{S}
\subset\mathcal{M}$. Then the following are true:
\begin{enumerate}
\item $I^{+}(x)$ is open in $\mathcal{M}$.

\item $I^{+}(\mathcal{S})$ is open in $\mathcal{M}$.

\item $I^{+}(\mathcal{S})=I^{+}(\overline{\mathcal{S}})$.

\item $I^{+}(\mathcal{S})=I^{+}\left( I^{+}(\mathcal{S})\right)\subset
J^{+}(\mathcal{S})=J^{+}\left( J^{+}(\mathcal{S})\right)$.
\end{enumerate}
\end{proposition}
\begin{proof}
To show that $I^{+}(x)$ is open in $\mathcal{M}$, take a $y\in
I^{+}(x)$. Then there exists a trip $\gamma$ from $x$ to $y$.
Consider a simple region $\mathcal{N}$ containing $\ y$ and choose
a point $z$ in $\mathcal{N}$ which lies on the terminal segment of
(the timelike, future oriented geodesic) $\gamma$. The initial
velocity of the (timelike, future--oriented) geodesic $\mu=zy$ is
$\mu' (0)=\exp_{z}^{-1}(y)$ and so it belongs to the open set
$\mathcal{Q}\subset\exp_{z}^{-1} (\mathcal{N})$ consisting of all
timelike, future--pointing vectors of
$\exp_{z}^{-1}(\mathcal{N}).$ Since $\exp_{z}$ is a homeomorphism,
the set $\exp_{z}(\mathcal{Q})$ is open in $\mathcal{M}$ and
obviously contains $y$ $(=\exp_{z}(\mu' (0)).$ By definition,
$\exp_{z}(\mathcal{Q})$ contains all points that can be joined to
$z$ by a timelike, future oriented geodesic and hence
$\exp_{z}(\mathcal{Q})\subset I^{+}(z).$ From the transitivity
property of $\ll$ we have that $I^{+}(z)\subset I^{+}(x)$ and the
result follows.

The second claim is obvious since $I^{+}(\mathcal{S})$ is an
arbitrary union of the open sets $I^{+}(x)$, $x\in\mathcal{M}$.

The direction $I^{+}(\mathcal{S})\subset
I^{+}(\overline{\mathcal{S}})$ in the third claim is obvious. To
prove the opposite inclusion take a $x\in\overline{\mathcal{S}}$
and suppose that $y\in I^{+}(x).$ Then $x\in I^{-}(y)$ and if
$x\in\partial\mathcal{S}$ (otherwise the result is obvious) then,
since $I^{-}(y)$ is open and every neighborhood of $x$ contains a
point in $\mathcal{S}$, it follows that there exists a
neighborhood $\mathcal{A}$ of $x$ in $I^{-}(y)$ containing a point
$z\in\mathcal{S}$. Hence $z\in I^{-}(y)$ and so $y\in I^{+}(z)$,
that is $y\in I^{+}(\mathcal{S}).$

Finally the last claim is shown as follows. Since $x\ll y$ implies
$x\prec y$ we have $I^{+}(\mathcal{S})\subset J^{+}(\mathcal{S})$
and we need to prove only the first and third equalities. For the
first one, if $x\in\mathcal{S}$ and $y\in I^{+}(x)$,  then the
transitivity property of $\ll$ implies the existence of a $z\in
I^{+}(x)$ such that $x\ll z\ll y$, or $y\in I^{+}(I^{+}(x))$ and
so $I^{+}(\mathcal{S})\subset I^{+}\left(
I^{+}(\mathcal{S})\right).$ If $y\in I^{+}\left(
I^{+}(\mathcal{S})\right)$ then there exists an $x\in\mathcal{S}$
and a $z\in I^{+}(\mathcal{S})$ such that $x\ll z\ll y.$ Thus
$x\ll y$ and hence $y\in I^{+}(\mathcal{S}).$ The last equality is
similar with $\prec$ replacing $\ll.$
\end{proof}

Some further important relations between chronology and causality
are given in the following
\begin{proposition}\label{further properties}
Let $x,y,z$ be points in $\mathcal{M}$. Then the following hold:
\begin{enumerate}
\item $x\ll y,\;y\prec z$ implies $x\ll z;$
\quad $x\prec y,\;y\ll z$ implies $x\ll z.$
\item  If $\mu$ is a null geodesic from $x$ to $y$ and $\nu$ is a null
geodesic from $y$ to $z$ then either $x\ll z$ or $\mu\cup\nu$ is a
single null geodesic from $x$ to $z$.
\item  If $x\prec y$ but $x\not\ll y$ (sometimes these conditions
are written $x\rightarrow y$ and
are called \emph{a horismos}), then there is a null geodesic from
$x$ to $y$.
\end{enumerate}
\end{proposition}
\begin{proof}
For the first suppose without loss of generality  that $x\ll y$,
$y\prec z$ and consider the trip $\mu=xy$ and the causal trip
$\nu=yz.$ Since $\nu$ is compact there is a finite number of
simple regions $\mathcal{N}_{1},...,\mathcal{N}_{r}$ that cover
$\nu.$ Set $p_{0}=y\in\mathcal{N}_{i_{0} }$ say. Then we choose:
The future endpoint of the connected component of $\nu$ in
$\overline{\mathcal{N}}_{i_{0}}$ from $p_{0},$ call it $p_{1},$
and also a point in $\mathcal{N}_{i_{0}}$, call it $q_{1}$, on the
final segment of $\mu$ different from $p_{0}.$ Then $q_{1}p_{1}$
is future-timelike. Then either $p_{1}=z$ in which case we are
done, or $p_{1}$ is not in $\mathcal{N}_{i_{0}}.$ In the latter
case suppose that $p_{1}$ is in some $\mathcal{N}_{i_{1}}.$ Then
we choose: The future endpoint of the connected component of $\nu$
in $\overline{\mathcal{N}}_{i_{1}}$ from $p_{1},$ call it $p_{2},$
and also a point in $\mathcal{N}_{i_{1}}$, call it $q_{2},$ on the
segment $q_{1}p_{1}$ different from $p_{1}.$ Then $q_{2}p_{2}$ is
future-timelike. Thus either $p_{2}=z$ in which case we are done,
or we can repeat the argument. The process will be terminated in a
finite number of steps since there is a finite number of connected
components of $\nu$ in $\overline{\mathcal{N}}_{i}.$

For the second claim suppose that $\mu\cup\nu$ is not a single
(null) geodesic then this implies that there is a `joint' at $y$
meaning that the direction of $\mu$ at $y$ is different from the
direction of $\nu$ at $y$. If we choose in a small neighborhood of
$y$, a point $y_{1}\in\mu$ and a point $y_{2}\in\nu$ then
$y_{1}y_{2}$ is a timelike geodesic and so we have $x\ll y_{1}\ll
y_{2}\prec y$, that is $x\ll y$ by 1.

The last result is shown as follows. If the causal trip $xy$
contains a timelike segment, then repeated application of (1)
gives $x\ll y.$ If on the other hand all segments of $xy$ are
null, then by (2), $x\ll y$ unless $xy$ is a single null geodesic.
This completes the proof of the last claim and that of the
Proposition.
\end{proof}

\begin{example}
The Einstein cylinder provides a counterexample of the converse of
(3) in Proposition \ref{further properties}. In this example,
there exist two points $x,y$ that can be joined by a null geodesic
between and also $x\ll y.$ Also there is another pair of points
$x,z$ in which $x$ and $z$ can be joined by two null geodesics and
$x\not\ll z.$
\end{example}

For reasons that will be clear later we define a \emph{causal
curve} to be a curve $\mu$ with the property that for all
$x,y\in\mu$ and for every open set $\mathcal{A}$ containing the
portion of $\mu$ from $x$ to $y,$ there is a causal trip from $x$
to $y$ (or from $y$ to $x$) contained entirely in $\mathcal{A}$.

Now fix a set $\mathcal{S\subset M}.$ We shall use the terminology
\emph{future set} for the chronological future
$I^{+}(\mathcal{S})$ and \emph{past set} for the chronological
past $I^{-}(\mathcal{S})$ of $\mathcal{S}.$ From Proposition
\ref{f-t} it follows that a set $\mathcal{F}$ is a future
(respectively past) set if and only if $\mathcal{F}=I^{+}
(\mathcal{F})$ (respectively $\mathcal{F}=I^{-}(\mathcal{F})$).
Future (and past) sets are obviously open sets, and if
$x\in\mathcal{F}$ and $x\ll y$ then also $y\in\mathcal{F}$.
Examples of future sets include the chronological and causal
futures of any set.

We call a set $\mathcal{S}$ \emph{achronal} if there are no two
points in $\mathcal{S}$ with a timelike separation, that is no two
points are chronologically related: if $x,y\in\mathcal{S}$, then
$x\not\ll y$. Examples of achronal sets in Minkowski space include
the future nullcone at the origin and the spacelike plane $t=z$.

Consider next the boundary $\partial$ of the future $I^+$ of a set
$\mathcal{S}$ (that is the boundary of a future set) in the
spacetime $\mathcal{M}$.
\begin{definition}[Achronal boundary]
A set $\mathcal{B}\subset\mathcal{M}$ is called an
\textbf{achronal boundary} if it is the boundary of a future set,
that is if
\begin{equation}\label{achr bound}
 \mathcal{B}=\partial I^{+}(\mathcal{S}).
\end{equation}
\end{definition}
Hence an achronal boundary $\mathcal{B}$ is an achronal set for no
two points in the boundary of a future set can be chronologically
related
($I^{+}(\overline{\mathcal{F}})\cap\partial\mathcal{F}=\varnothing$,
where the closure of $\mathcal{F}$ is defined as
$\overline{\mathcal{F}}=\mathcal{F}\cup\partial\mathcal{F}$). It
is also not difficult to establish that $B$ cannot be spacelike
either (apart from possibly at $\mathcal{S}$ itself) and therefore
it must be a null set. Achronal boundaries are topological (ie.,
not necessarily smooth) 3-manifolds and are generated  (i.e., made
out) from null geodesics. The important properties of these null
geodesic generators of achronal boundaries are summarized in the
following result which we give without proof.
\begin{theorem}
Let $\mathcal{S}\subset\mathcal{M}$ and $B=\partial
I^{+}(\mathcal{S})$. If
$x\in\mathcal{B}\setminus\bar{\mathcal{S}}$ is a future endpoint
of a null geodesic $\mu\in\mathcal{B}$ then $\mu$ is either
past-endless or has a past endpoint on $\bar{\mathcal{S}}$. Also
every future extension of $\mu$ must leave $\mathcal{B}$ and enter
$I^{+}(\mathcal{S})$.
\end{theorem}
This basically says that every null geodesic generator has a
future endpoint in the achronal boundary and if it intersects
another generator it will have to leave the boundary and enter
into the interior of the future. On the other hand, null geodesic
generators are either past endless or can have past endpoints only
on $\mathcal{S}$.

Let $\mathcal{S}$ be achronal. We define another achronal set the
\emph{edge of} $\mathcal{S}$, $\textrm{edge}(\mathcal{S})$, which
is the set of points $x\in\bar{\mathcal{S}}$ such that every
neighborhood $\mathcal{Q}$ of $x$ contains points $p\in I^{-}(x)$
and $q\in I^{+}(x)$, which can be joined by a trip in
$\mathcal{Q}$ that does not intersect $\mathcal{S}$. If
$\textrm{edge}(\mathcal{S})=\varnothing$, then $\mathcal{S}$ is
called \emph{edgeless}. Every edgeless set must be closed.

There are three important constructions closely related to the
notion of achronality, namely the domain of dependence, the Cauchy
horizon and the Cauchy surface. We discuss each one of these in
turn.

Let $\mathcal{S}$ be an achronal, closed subset of $\mathcal{M}$.
\begin{definition}[Domains of dependence]
The \textbf{domain of dependence} (or Cauchy development) of
$\mathcal{S}$ is the set,
\begin{equation}\label{dom dep}
  D(\mathcal{S})=\Bigl\{ p\in\mathcal{M}:\;\textrm{every
  endless trip through}\; p\; \textrm{meets} \;\mathcal{S}\Bigr\}.
\end{equation}
The \textbf{future domain of dependence} of $\mathcal{S}$ is the
set,
\begin{equation}\label{dom dep+}
  D^{+}(\mathcal{S})=\Bigl\{ p\in\mathcal{M}:\;\textrm{every
  past-endless trip through}\; p\; \textrm{meets} \;\mathcal{S}\Bigr\}.
\end{equation}
The \textbf{past domain of dependence} of $\mathcal{S}$ is the
set,
\begin{equation}\label{dom dep-}
  D^{-}(\mathcal{S})=\Bigl\{ p\in\mathcal{M}:\;\textrm{every
  future-endless trip through}\; p\; \textrm{meets} \;\mathcal{S}\Bigr\}.
\end{equation}
\end{definition}
It follows that $D(\mathcal{S})=D^{+}(\mathcal{S})\cup
D^{-}(\mathcal{S})$ and that $\mathcal{S}$ is contained in
$D(\mathcal{S})$. Thus $D^{+}(\mathcal{S})$ is a closed set and
denotes the region that can be predicted from knowledge of data on
$\mathcal{S}$. What are the future limits of that region? In other
words, what is the set of those points $p\in D^{+}(\mathcal{S})$
such that events in $I^{+}(p)$ do not belong to
$D^{+}(\mathcal{S})$? These questions lead to another achronal,
closed set -- the future boundary of the future domain of
dependence: The future Cauchy horizon.
\begin{definition}[Cauchy horizon]
The \textbf{future, past and total Cauchy horizon} of
$\mathcal{S}$ is defined as respectively,
\begin{equation}\label{h+}
  H^{+}(\mathcal{S})=\Bigl\{ p\in\mathcal{M}:\;p\in
  D^{+}(\mathcal{S})\;\textrm{but}\; I^{+}(p)\cap
  D^{+}(\mathcal{S})=\varnothing\Bigr\} ,
\end{equation}
\begin{equation}\label{h-}
  H^{-}(\mathcal{S})=\Bigl\{ p\in\mathcal{M}:\;p\in
  D^{-}(\mathcal{S})\;\textrm{but}\; I^{-}(p)\cap
  D^{-}(\mathcal{S})=\varnothing\Bigr\} ,
\end{equation}
\begin{equation}\label{h}
  H(\mathcal{S})=H^{+}(\mathcal{S})\cup  H^{-}(\mathcal{S}).
\end{equation}
or equivalently,
\begin{equation}\label{equiv h+-}
   H^{\pm}(\mathcal{S})=D^{\pm}(\mathcal{S})\setminus
   I^{\mp}\bigl(D^{\pm}(\mathcal{S})\bigr) .
\end{equation}
\end{definition}
The future Cauchy horizon $H^{+}(\mathcal{S})$ is another example
of a achronal, closed set and it holds that
$H(\mathcal{S})=\partial D(\mathcal{S})$. Finally we have:
\begin{definition}[Cauchy surface]
A \textbf{Cauchy surface for} $\mathcal{M}$ is an achronal set
$\mathcal{S}$ such that,
\begin{equation}\label{cauchy surface}
D(\mathcal{S})=\mathcal{M}.
\end{equation}
\end{definition}
To have a Cauchy surface $\mathcal{S}$ in a spacetime
$\mathcal{M}$ is a statement for both $\mathcal{S}$ and
$\mathcal{M}$. Intuitively speaking, if $\mathcal{M}$ has a Cauchy
surface then initial data on $\mathcal{S}$ determine the entire
past and future evolution of $\mathcal{M}$. The existence of a
Cauchy surface is a global condition to impose on a spacetime, and
it can happen that a surface $\mathcal{S}$ may appear to be a
Cauchy surface for a spacetime $\mathcal{M}$ during an early stage
in the evolution, but later $\mathcal{M}$ may develop in such a
way so that no Cauchy surface can be admitted.

To proceed further we need to impose some \emph{global} causality
assumption on our spacetime. There is a number of such assumptions
on the market the most important and also the most restrictive of
all being the fundamental concept of global hyperbolicity, or
\emph{hyperbolicit\'{e} globale} if one wishes to be historically
just!
\begin{definition}[\emph{Global hyperbolicity}]
A spacetime $\mathcal{M}$ is called \textbf{globally
hyperbolic} if it satisfies the following two conditions:
\begin{description}
  \item[Strong causality] It contains no closed or almost closed
  trips
  \item[Compact diamond-shapes] For any two points $p,q$,  the
  intersection $I^{+}(p)\cap I^{-}(q)$ has compact closure
  (or equivalently, the diamond-shaped sets $J^{+}(p)\cap
  J^{-}(q)$ are compact)
\end{description}
\end{definition}
The compactness of the sets $J^{+}(p)\cap J^{-}(q)$ means that
these diamond-shaped sets do not contain points at infinity or
singular points, that is points that can be regarded as belonging
to spacetime's `edge'. This second condition requires intuitively
speaking that in the region between any pair of points in
spacetime there are no asymptotic regions or holes or
singularities. If on the other hand strong causality is violated
in a spacetime this must be due to some \emph{global} feature. In
such a case, trips or causal trips starting near $p$ will return
to points near $p$ without necessarily being closed. Thus global
hyperbolicity is primarily a `cosmological' condition. It can be
shown that global hyperbolicity is equivalent to the existence of
a Cauchy surface and is a stable property with respect to
sufficiently small perturbations of the metric. We refer the
interested reader to the excellent exposition of these basic
causality properties given in \cite{ge70}.

We now give the last definition we need from causal structure
theory.
\begin{definition}[Trapped]
A \textbf{future-trapped set} is an achronal, closed set
$\mathcal{S}\subset\mathcal{M}$ for which the set,
\begin{equation}\label{f trapped}
E^{+}(\mathcal{S})=J^{+}(\mathcal{S})\setminus I^{+}(\mathcal{S}),
\end{equation}
called the future horismos of $\mathcal{S}$, is compact.
\end{definition}
It follows that since $\mathcal{S}\subset E^{+}(\mathcal{S})$, any
future-trapped set is itself compact. A very special example of a
future-trapped set is an achronal, closed, spacelike hypersurface.
The dual definition, a past-trapped set, is obvious.

\section{Globalization and singularity theorems}\label{globalization}
The results of the Section \ref{conjugate points and geodesic
congruences} are local in character. They are concerned with
conditions for the existence of length-maximizing curves in
spacetime and obstructions to such curves. These conditions, we
showed, are about the absence or existence of conjugate points to
points or spacelike hypersurfaces in spacetime. However, the
question of interest to us has to do with the structure of
spacetime \emph{globally}, that is on the whole, and it is unclear
how the local results obtained so far concerning the behaviour of
congruences of curves \emph{in} spacetime can somehow be elevated
to reveal information about the global behaviour of the spacetime
\emph{itself}. It is clear we need some method to \emph{globalize}
them. It is the purpose of the present Section to discuss this
problem.

Consider the set $\mathcal{K}$ of points in which $\mathcal{M}$ is
strongly causal (no closed or almost closed trips), a compact
subset $\mathcal{C}$ of $\mathcal{K}$ and two closed subsets
$\mathcal{A}$ and $\mathcal{B}$ of $\mathcal{C}$. Denote by
$\mathcal{D}$ the set of all trips in $\mathcal{K}$, by
$\mathcal{E}$ the set of all causal trips in $\mathcal{K}$ and by
$\mathcal{F}$ the set of all causal curves in $\mathcal{K}$. It is
clear that $\mathcal{D}\subset\mathcal{E}\subset\mathcal{F}$. We
are interested in the sets of all causal curves in $\mathcal{C}$
from a point of $\mathcal{A}$ to a point of $\mathcal{B}$ which we
denote by $C_{\mathcal{C}}\bigl(\mathcal{A},\mathcal{B}\bigr)$
with $C\bigl(\mathcal{A},\mathcal{B}\bigr)$ meaning the set of all
causal curves in $\mathcal{K}$.

The basic idea can be summarized without proofs by the following
\emph{globalization procedure}:
\begin{enumerate}
  \item[GP1] Put a suitable topology on
  $C_{\mathcal{C}}\bigl(\mathcal{A},\mathcal{B}\bigr)$ so that
  $C_{\mathcal{C}}\bigl(\mathcal{A},\mathcal{B}\bigr)$ becomes a
  compact set
  \item[GP2] Define a suitable length function $L: C_{\mathcal{C}}\bigl(\mathcal{A},
  \mathcal{B}\bigr)\longrightarrow\mathbb{R}$
  on this set and show it is upper semicontinuous
  \item[GP3] Compactness on
  $C_{\mathcal{C}}\bigl(\mathcal{A},\mathcal{B}\bigr)$
  implies that $L$ attains a maximum value on
  $C_{\mathcal{C}}\bigl(\mathcal{A},\mathcal{B}\bigr)$
  \item[GP4] This maximum is always attained in globally hyperbolic spacetimes and
  can be arranged to be on a geodesic without conjugate points.
\end{enumerate}
In this way all arguments of  Section \ref{conjugate points and
geodesic congruences} can be globalized and we have extended the
definition of the length of a smooth curve to that of a
\emph{continuous} curve\footnote{By a technical argument which we
omit, we can rule out the possibility that a continuous, nonsmooth
curve exists which has length greater or equal to that of any
geodesic. In essence one shows that a continuous, nonsmooth curve
connecting any two points cannot be length maximizing, for if it
fails to be a geodesic at a point we can deform it, in a convex
normal neighborhood of that point, to obtain a curve of greater
length.}.

We are now in a position to prove the simplest singularity
theorem.
\begin{theorem}\label{first thm}
Let $(\mathcal{M},g_{ab})$ be a spacetime such that the following
conditions hold:
\begin{enumerate}
  \item $(\mathcal{M},g_{ab})$ is globally hyperbolic
  \item $R_{ab}\,V^{a}\,V^{b}\geq 0$ for all timelike vectorfields
  $V^{a}$
  \item There exists a smooth spacelike Cauchy surface $\Sigma$ such that
  the expansion $\theta$ of the past-directed geodesic congruence
  orthogonal to $\Sigma$ satisfies $\theta\leq C< 0$, where $C$ is a constant,
  everywhere on $\Sigma$.
\end{enumerate}
Then no past-directed timelike curve from $\Sigma$ can have length
greater than $3/|C|$, that is all past-directed timelike geodesics
are incomplete.
\end{theorem}
\begin{proof}
The proof goes by \emph{reductio ad absurdum}. Suppose there is a
past-directed timelike curve $\mu$ and $p$ a point on $\mu$ lying
beyond length $3/|C|$ from $\Sigma$. By (GP4), there exists a
curve $\gamma$ of maximum length in $C(\Sigma ,p)$ which obviously
must have length greater than $3/|C|$ and $\gamma$ must be a
geodesic with no conjugate points between $\Sigma$ and $p$.
However, this contradicts Theorem \ref{conj point thm} which
predicts the existence of a conjugate point between $\Sigma$ and
$p$. Therefore the curve $\mu$ cannot exist.
\end{proof}

The following is (a corollary of) a general singularity theorem
which combines both past and future singularities. Its proof can
be found in pages 266-270 of \cite{ha-el73}.
\begin{theorem}\label{thm2}
$(\mathcal{M},g_{ab})$ cannot be timelike and null geodesically
complete if:
\begin{enumerate}
  \item $R_{ab}\,V^{a}\,V^{b}\geq 0$ for all timelike and null vectorfields
  $V^{a}$
  \item The generic conditions hold for all timelike and null
  vectorfields
  \item There are no closed timelike curves (the \emph{chronology
  condition})
  \item $(\mathcal{M},g_{ab})$ possesses at least one of the following:
  (a) A compact, achronal set without edge,  (b) a closed trapped surface, (c) a point $p$ in
  $\mathcal{M}$ such that the expansion of every past (or every future) null geodesic
  congruence emanating from $p$ becomes negative along each
  geodesic of this congruence (i.e., the null geodesics from $p$
  refocus).
\end{enumerate}
\end{theorem}

\section{Cosmological applications}
The simplest singularity theorem, Theorem \ref{first thm} (and in
fact all such theorems about past-incomplete spacetimes), has
profound implications in cosmology, for it predicts the existence
of a cosmological singularity in the past a finite time ago in the
form of timelike geodesic incompleteness for a universe which is
globally hyperbolic and everywhere expanding at one instant of
time. We now explain how this arises.

In order to apply the purely mathematical results about geodesic
incompleteness discussed above to cosmology in a meaningful way,
we have to connect somehow the geometry of the situation to the
behaviour of matter in the real universe. As we discussed in the
Introduction, we shall follow Einstein and postulate an
interaction of spacetime geometry and the distribution of matter
in the universe through the Einstein field equations,
\begin{equation}\label{efe}
    R_{ab}-\frac{1}{2}\;g_{ab}R=T_{ab}\;.
\end{equation}
When this is assumed, the purely geometric assumptions and results
present in the singularity theorems above acquire immediate
physical meanings. We comment on the physical ramifications of the
singularity theorems here very briefly, referring those interested
to the original papers \cite{p1}-\cite{hp1}.

Let us start with the convergence conditions (Assumption 2 of
Theorem \ref{first thm} or Assumption 2 of Theorem \ref{thm2}). If
we write the Einstein equations in the equivalent form,
\begin{equation}\label{efe1}
    R_{ab}=T_{ab}-\frac{1}{2}\;g_{ab}T\;,
\end{equation}
the timelike convergence condition from Definition \ref{cc} yields
restrictions on the matter content through the energy-momentum
tensor.
\begin{definition}[Energy conditions]
The energy-momentum tensor $T_{ab}$ satisfies the \textbf{strong
energy condition} if,
\begin{equation}\label{sec}
T_{ab}V^{a}V^{b}-\frac{1}{2}\;TV^{a}V_{a}\geq 0\quad\textrm{for
all timelike vectorfields}\; V^{a}.
\end{equation}
It satisfies the \textbf{weak energy condition} if,
\begin{equation}\label{wec}
T_{ab}V^{a}V^{b}\geq 0\quad\textrm{for all timelike
vectorfields}\; V^{a}.
\end{equation}
\end{definition}
The weak energy condition is a weaker requirement on the
energy-momentum tensor than the strong energy condition. By
continuity, the former is also true for all null vectorfields.

We see that the singularity theorems place restrictions on the
matterfields in the universe in the form of energy conditions on
the energy-momentum tensor, independently of the detailed form of
the matterfields. Also as we said in the Introduction, they do not
assume any splitting in the form of the matter tensor. For
example, since for an observer whose worldlines have tangent
vectorfield $V^{a}$ we have,
\begin{equation}\label{energy density}
  T_{ab}V^{a}V^{b}=\textrm{energy density}\;,
\end{equation}
it  follows that the weak energy condition means that the energy
density as measured by any observer is non-negative, which seems a
very reasonable assumption on matter.

The strong energy condition is stricter than the weak energy
condition but it is also reasonable on macroscopic scales since,
through the Einstein equations, it is the corresponding inequality
to the timelike convergence condition and so it means that
`gravitation is always attractive' in the sense that neighboring
geodesics accelerate on the average towards each other. Only a
positive cosmological constant $\Lambda >0$ can induce a cosmic
repulsion thus preventing gravity from being always attractive.
Thus in all theories which in the Einstein frame (see Section
\ref{cosmologies}) become like general relativity without a
positive cosmological constant, and provided gravity remains
attractive and the other conditions of the theorems hold, the
singularity theorems apply.

The existence of closed timelike curves leads to severe
difficulties of interpretation. For the simplest wave equation
$u_{tt}-u_{xx}=0$ on the $(x,t)$-torus, the only solution with
$(t,x)$ identified with $(t+n,x+m\pi )$ is the $u
=\textrm{const.}$ solution. So, it is not necessary for the
singularity problem to consider closed timelike curves.

The generic conditions can fail along a geodesic only if we
consider very special models (for more information, see
\cite{hp1}, page 540). These conditions are therefore very general
too.

We also see that the assumption of global hyperbolicity of Theorem
\ref{first thm} is absent from the general Theorem \ref{thm2}, and
assumption 4 of this Theorem contains a much weaker version of
assumption 3 of Theorem \ref{first thm} namely, the universe is
not assumed to be expanding everywhere. However, the results of
this general singularity theorem are somewhat weaker as no
information is given as to whether the singularity is in the
future or past. One expects, however, that when a closed trapped
surface is present the singularity is in the future and when the
past nullcone starts reconverging the singularity is in the past.

We close this chapter with one final remark about the
singularities discussed above. In general relativity singularities
are described as timelike and null geodesic incompleteness and are
shown to exist as a result of the singularity theorems proved
above. These theorems use the Jacobi equation for the geodesic
variation vectorfield, the timelike convergence (equivalently the
energy) condition and some topological condition that comes as a
result of a causality requirement. They do \emph{not} use the
Einstein equation (\ref{efe}). On the other hand, studies of the
Cauchy and global existence problems for Eq. (\ref{efe}) (cf.
\cite{ch-kl93,ch-mo01}) indicate that the general solution of Eq.
(\ref{efe}) exists for an infinite proper time and lead us to
suspect or expect that there would be no \emph{dynamical}
singularities for the Einstein field equations but these only
appear when we impose some unphysical or nongeneric symmetry
assumption (for example that the universe is described by a
Friedmann or Bianchi metric). What is the relation between the
apparent absence of any dynamical singularities of the Einstein
equations  and the general relativistic singularities in the
geometric Hawking-Penrose sense of g-incompleteness via the Jacobi
equation? We refer the reader interested in such questions to the
very recent work \cite{ch-co02}.

The singularity theorems discussed in these notes are important
results for cosmology and point to an aspect of cosmology which
should not be forgotten namely, the application of rigorous
mathematical techniques for the study of the universe is not a
luxury. It may sometimes lead to profound changes of viewpoint for
the approach and possible resolution of current and future
cosmological issues. In many respects it reminds us of the
well-known inscription in Plato's Academy:
\begin{center}
\textgreek{ΜΗΔΕΙΣ ΑΓΕΩΜΕΤΡΗΤΟΣ ΕΙΣΙΤΩ}.
\end{center}

\end{document}